\documentclass[fleqn, adp, a4paper]{w-art}
\usepackage{graphics,graphicx}
\usepackage{amsmath}
\usepackage{amssymb}
\usepackage[T1]{fontenc}
\usepackage{psfrag}
\usepackage{dcolumn}
\usepackage{color}
\usepackage{amsfonts}
\usepackage{dsfont}
\usepackage{setspace}

\usepackage[fleqn]{amsmath}
\def\vct#1{\mathbf{#1}}
\newcommand{\be}{\begin{equation}}
\newcommand{\ee}{\end{equation}}
\newcommand{\bea}{\begin{eqnarray}}
\newcommand{\eea}{\end{eqnarray}}
\newcommand{\bs}{\begin{subequations}}
\newcommand{\es}{\end{subequations}}

\def\no{\nonumber \\}

\def\nunit{n_{\rm{12}}}
\newcommand{\vnunit}{\vct{\nunit}}
\newcommand{\vvelocity}{\vct{v}}

\newcommand{\Mean}{{\cal M}}
\newcommand{\MeMo}{{\cal N}}
\newcommand{\EccAno}{u}

\newcommand{\peri}{{\cal K}}
\newcommand{\periEffective}{{\cal K}}

\newcommand{\cinv}{c^{-1}}

\newcommand{\Max}{{\mathfrak M}}

\newcommand{\vAng}{\vct{h}}

\newcommand{\OTS}{\,{\mathcal W}_{e_t}}

\newcommand{\En}{|E|}

\newcommand{\magnetic}{m}
\newcommand{\vKE}[1]{G_{#1}}

\newcommand{\CKE}[2]{{\cal \bar \gamma}^{#1}_{#2}}
\newcommand{\SKE}[2]{{\cal \bar \sigma}^{#1}_{#2}}
\newcommand{\CKEN}[2]{ {	\overset{0}{\cal \bar \gamma}{^{#1}_{#2}}	}}
\newcommand{\SKEN}[2]{ {	\overset{0}{\cal \bar \sigma}{^{#1}_{#2}}	}}
\newcommand{\CKETwoPN}[2]{ {	\overset{4}{\cal \bar \gamma}{^{#1}_{#2}}	}}
\newcommand{\SKETwoPN}[2]{ {	\overset{4}{\cal \bar \sigma}{^{#1}_{#2}}	}}

\newcommand{\SAKE}[2]{{\cal S}^{(#1)}_{#2}} %\colblu
\newcommand{\AKE}[2]{{\cal A}^{(#1)}_{#2}} %\colblu

\newcommand{\ProdSS}[1]{P^{SS, [n]}_{#1}}
\newcommand{\ProdCS}[1]{P^{CS, [n]}_{#1}}
\newcommand{\ProdCC}[1]{P^{CC, [n]}_{#1}}

\newcommand{\funcAnq}[2]{{\cal  F}_{[#1 #2]}(u)}
\newcommand{\funcSinAnq}[2]{{\cal F}_{S[#1 #2]}(u)}

\newcommand{\salpha}{{\tilde \alpha}}
\newcommand{\sbeta}{{\tilde \beta}}

\newcommand{\funcSinAKEc}[1]{	({\cal \bar S}_c)^{[#1]}} %\colviol%{	({\cal SA}c)^{[#1]}}
\newcommand{\funcSinAKEs}[1]{	({\cal \bar S}_s)^{[#1]}} %\colviol%{	{\cal SA}s)^{[#1]}}

\newcommand{\funcAKEc}[1]{	({\cal \bar A}_c)^{[#1]}} %\colviol%{	({\cal A}c)^{[#1]}}
\newcommand{\funcAKEs}[1]{	({\cal \bar A}_s)^{[#1]}} %\colviol%{	({\cal A}s)^{[#1]}}

\newcommand{\TensorProdttpp}{\left( \hat{\theta} \hat{\theta} -\hat{\phi} \hat{\phi} \right)}
\newcommand{\TensorProdtppt}{ \left( \hat{\theta} \hat{\phi} + \hat{\phi} \hat{\theta} \right)}

\allowdisplaybreaks

\begin{document}
\setlength\mathindent{0pt}
%\title[Frequency domain 1\pN\ accurate gravitational wave forms from eccentric compact binaries]
\title[Full analytic frequency-domain GW forms]{Full-analytic frequency-domain gravitational wave forms from eccentric compact binaries to 2PN accuracy}
% \author[M. Tessmer]{Manuel Tessmer\inst{1}%
%   \footnote{Corresponding author: Manuel Tessmer\quad E-mail:~\textsf{m.tessmer@uni-jena.de}, 
% }}
\author[M. Tessmer]{Manuel Tessmer \footnote{Manuel Tessmer: e-mail: {\sf m.tessmer@uni-jena.de}, Phone: +49\,(0)3641\,947108}\inst{1}}

\author[G. Sch\"afer]{Gerhard Sch\"afer\inst{1}}
\address[\inst{1}]{Theoretisch-Physikalisches Institut, Friedrich-Schiller-Universit\"at Jena, Max-Wien-Platz 1, 07743 Jena, Germany}
\date{\today}
%\keywords{Gravitational waves, }

\date{\today}

\begin{abstract}
The article provides full-analytic gravitational
wave (GW) forms
for eccentric nonspinning compact binaries
of arbitrary mass ratio in the time Fourier
domain. The semi-analytical
property of recent descriptions, i.e. the demand
of inverting the higher-order Kepler equation
numerically but keeping all other
computations analytic, is avoided for the first time.

The article is a completion of  a previous
one
{(Tessmer and Sch\"afer, {Phys. Rev. D} \textbf{82}, 124064 (2010))}
to
second post-Newtonian (2PN) order in the harmonic GW
amplitude and conservative orbital dynamics.
A fully analytical inversion formula
of the Kepler equation in harmonic coordinates
is provided, as well as the analytic time Fourier expansion
of trigonometric functions of the eccentric anomaly
in terms of sines and cosines of the mean anomaly.
Tail terms are not considered.
\end{abstract}

\maketitle
% \input{HTTFou2PNIntro}
% \input{HTTFou2PNOrbit}
% \input{HTTFou2PNMomenta}
% \input{HTTFou2PNWaveForm}
% \input{HTTFou2PNBasicIngredients}
% %\input{HttFou2PNMultipoleDecomp}
% \input{HTTFou2PNResults}
% \input{HTTFou2PNConclusion}
% \input{HTTFou2PNAck}
% \input{HTTFou2PNAppendix}
% \input{HTTFou2PNTensorSphHarmonics}
% \input{HTTFou2PNAppendixResults}

%=======================================================
\section{Introduction}

In our  previous article \cite{Tessmer:Schafer:2010}
-- it will be  called ``paper 1'' from now onwards --
we provided
the Fourier domain gravitational wave (GW) forms including the
first post-Newtonian (1PN) corrections
to the leading order quadrupolar contribution and 1PN corrections
to the Newtonian accurate orbital dynamics in terms of tensor
spherical harmonics. In this article, we complete the amplitude
to 2PN order in harmonic coordinates and we include 2PN orbital
dynamics as well.
We regard nonspinning compact binaries with arbitrary mass ratio
in quasi-elliptical orbits, where the orbital dynamics is well-known.
Earlier works provided the solution to the evolution equations in
terms of a Keplerian parameterization,
perturbed by their post-Newtonian corrections. The result is the
so-called quasi-Keplerian parameterization (QKP), to be found,
e.g. in \cite{Damour:Deruelle:1985, 
Memmesheimer:Gopakumar:Schafer:2004, Damour:Gopakumar:Iyer:2004},
which will be one basic for this article.
2PN corrections to the GW amplitude in the limit of non-spinning binaries
are also dealt with in \cite{Gopakumar:Iyer:2002} where the wave forms
are provided with the help of a quasi-Keplerian parameterization,
and recently, including spin, in \cite{Cornish:Key:2010}
where the results are given in terms of coordinate velocities.
We like to improve the use of those wave forms as we provide them as
{\em pure series of harmonics in the time Fourier domain}, as this is
one step closer to data analysis investigations.
{Let us now briefly summarise why 2PN corrections and the inclusion of eccentric orbits are necessary.
Current data analysis investigation showed that, even for circular
inspirals, the inclusion of 2PN orbital dynamics is necessary
to guarantee a successful detection of the GW signal
\cite{Damour:Gopakumar:Iyer:2004, Gopakumar:Hannam:Husa:Brugmann:2008, Reisswig:Husa:Rezzolla:Dorband:Pollney:Seiler:2009}.
% To perform an eccentric GW analysis
% for LISA is demanded, because even orbits with eccentricity
% $e \approx 0.001$ can be distinguished from circular ones
% \cite{Key:Cornish:2010}.}
Authors of a recent publication \cite{AmaroSeoane:Eichhorn:Porter:Spurzem:2010}
investigated certain equal-mass black hole binaries in the millisecond GW frequency
range in rotating clusters and stated parameter estimation errors for the initial
eccentricity of $\sim 10^{-7}$ for LISA. Therefore we claim that to perform
an eccentric GW analysis for LISA is demanded.
}
The GW energy flux in terms of its harmonic constitutive parts,
was given to 1PN order, in the extreme mass-ratio regime, for the first time 
 -- to our best knowledge --  in \cite{Galtsov:Matiukhin:Petukhov:1980}.
First attempts, however, for a time Fourier domain (TFD)
of eccentric $h_+$ and $h_\times$, the two polarizations of the far-zone GW field,
have been made in Reference \cite{MorenoGarrido:Buitrago:Mediavilla:1994}
where the periastron advance has been incorporated by hand and the stationary
phase approximation has come into use, in \cite{MorenoGarrido:Mediavilla:Buitrago:1995}
and \cite{Pierro:Pinto:Spallicci:Laserra:Recano:2001}, where
a spectral analysis for steady-state binaries in the simple case of Newtonian motion
and amplitude has been performed.

Recently in { \cite{Yunes:Arun:Berti:Will:2009}}, 
the authors furnished Newtonian
accurate TFD wave forms as they incorporated the sine
and cosine functions of the eccentric anomaly as Fourier--Bessel series.
Taking as starting point $h_+$ and $h_\times$ themselves, having
evaluated all appearing scalars (such as, for example, the scalar product
of the basic vectors of the plane of the sky with the orbital velocity vector)
being evaluated explicitly, would lead to rather complicated
relations of rotation angles connecting the
orientation of the binary's orbital plane with respect to the position of the
observer (see, e.g. \cite{Tessmer:Hartung:Schafer:2010}, and -- including spin
precession -- \cite{Arun:Buonanno:Faye:Ochsner:2009}  for higher orders
of $h_+$ and $h_\times$) and this is not well-suited for a systematic
TFD representation at higher orders.

%Anyway, there is an outlet.
The problem of nonspinning compact binaries % is such one which
implies the conservation of the direction of the orbital angular
momentum $\vAng$, thus, it qualifies for a representation with the help of
tensor spherical harmonics \cite{Thorne:1980, Turner:Will:1978, Maggiore:2007}.
Reference \cite{Junker:Schafer:1992} provided 1PN accurate
spherical tensor components for nonspinning compact binaries in
quasi-elliptic orbits (and quasi-hyperbolic ones as well) that we are on to verify
in this article. Irrespective of the observer's orientation, the form
the GW signal can be expressed with the help of these spherical tensor
components and applied to give the
polarizations $h_+$ and $h_\times$ {\cite{Blanchet:Damour:Iyer:1995}}
afterwards with the help of constant rotations.

We provide an analytical expression of the eccentric anomaly $u$ in terms of
the mean anomaly $\Mean$, and do likewise for the sines and cosines of arbitrary
integer multiples of $u$ at full 2PN order. By means of this, we are able to
express the spherical tensor components in terms of infinite series in harmonics of
the elapsed time. To which order the infinite summations have to be driven will be investigated.
Similarly to Reference \cite{Gopakumar:Iyer:1997}, tail terms occurring through 2PN order
will not be discussed.

\subsection{Organisation of the paper}
The paper is organized in the following way.
Section \ref{Sec:orbit} collects the orbital elements
of the quasi-Keplerian presentation of the 2PN-accurate
orbital motion in harmonic coordinates.
Section \ref{Sec::Multipoles} summarizes the transverse-traceless (TT)
projection of the far-zone GW field up to 2PN order corrections to the
leading-order quadrupole field. The time derivatives of the source multipole
moments are computed using expressions for the particle accelerations known
from the literature, and the scalar components
are projected out with the help of TT tensors in Section \ref{Sec:WaveForm}.
The vital components of the Fourier decomposition, here at 2PN accuracy,
of the GW field are summarized in Section \ref{Sec:BasicIngredients} and
the Fourier domain multipoles, incorporating purely 2PN conservative dynamics,
are given structurally in its Subsection \ref{Sec::Results}.
The full explicit expressions can be
found in Appendix \ref{Appendix::Results}.
\newpage

\subsection{Notation}

An overview of the quantities which are in use in this paper is given in form of
a table of their associated abbreviations and scalings.

\begin{table}[!h]
\begin{tabular}{l  l  r c}
$\EccAno$:	&~	& \dotfill eccentric anomaly & \hspace{2cm}\\
$\Mean$:	&~	& \dotfill mean anomaly &\\
$\vct{h}$:
		&	& \dotfill orbital vectorial angular momentum, $h:=|\vct{h}|$ &\\
xPN:		&	& \dotfill x$^{\rm th}$ post Newtonian &\\
$r$		&	& \dotfill radial separation of the binary components&\\
$e_r$	&	& \dotfill ``radial eccentricity''	&\\
$e_t$	&	& \dotfill ``time eccentricity''	&\\
$e_\phi$	&	& \dotfill ``phase eccentricity'' &\\
$a_r$	&	& \dotfill semimajor axis	&\\
$\phi$	&	& \dotfill orbital phase &\\
$v$		&	& \dotfill true anomaly &\\
$\Phi$	&	& \dotfill  total phase elapsed from a periastron to periastron time &\\
$\MeMo$	&	& \dotfill mean motion, $\MeMo:=\frac{2\pi}{P}$ where $P$ is the time from periastron to periastron &\\
$\peri$	&	& \dotfill periastron advance parameter, $\peri:= \frac{\Phi-2\pi}{2\pi}$ &\\
$\eta$	&	& \dotfill symmetric mass ratio, $\eta:=\frac{m_1 \, m_2}{m^2}$, $m=m_1+m_2$ as the sum
				of the masses &\\
$\delta_m$
		&	& \dotfill the difference in the masses, $\delta_m:=m_1-m_2$ &\\
$\En$	&	& \dotfill absolute value of the scaled orbital binding energy &\\
$G$	&	& \dotfill Newtons gravitational constant &\\
$c$	&	& \dotfill speed of light 
% & \\ \hspace{\textwidth}
\end{tabular}
\end{table}
Multiple indices are used according to Thorne and Blanchet {\cite{Thorne:1980, Blanchet:Damour:1986} }. 
In the final form of our results, energies are scaled with
$\mu c^2$, separations with $\frac{G\,m}{c^2}$, momenta with $\mu c$  and times with $\frac{G\,m}{c^3}$, such that all quantities are dimensionless.
Lower index ``0'' labels the value of a quantity at time  $t=t_0$.
%=======================================================
\subsection*{Schematic prescription of our calculation}
% \colblu{Here we give a brief overview about what we do.
% The basis is the expansion of the GW field in terms of multiple time derivatives
% of multipole moments, see the subsequent sections. In these time derivatives, the
% accelerations will be eliminated, as they can be expressed in terms of velocities and positions only.
% The terms that are occurring here have free indices after contraction, as we are in the basis of STF tensors and the GW
% amplitudes are built via contractions with the line-of-sight vector.
% We can, however, change the basis and project them onto tensor spherical harmonics (TSH) and obtain the {\em scalar spherical components}
% which will be the central topic of our investigation.
% The distances on the one hand therein and the velocities on the other that are remaining after this operation\
% will be expressed in terms of the binding energy, the eccentricity and -- implicitly -- the elapsed time, as we use the
% QKP for compact binaries in harmonic coordinates.
% The terms turn out to be even and odd functions of $\EccAno$ and, thus, of $\Mean$. We will successively
% find decompositions in terms of harmonics in $\EccAno$ and those of $\Mean$ with {\em positive} frequencies.
% \\
% }
For convenience, we show a flow diagram of our computation. Preliminary results have been surrounded by boxes.\\
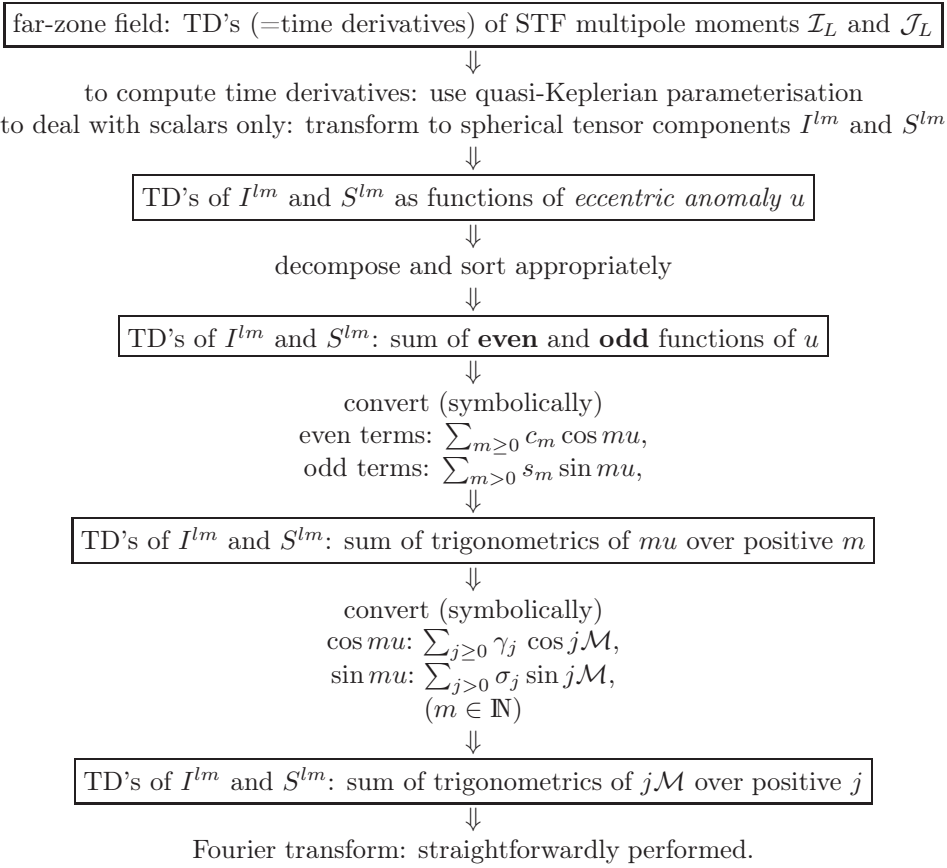
\vspace{1cm}
\begin{center}
\begin{tabular}{r  c  l}
					\multicolumn{3}{c}{ \fbox{far-zone field: TD's (=time derivatives) of STF multipole moments ${\cal I}_L$ and ${\cal J}_L$}}	\\
					&			$\Downarrow$							&	\\
% QKP 	$\Rightarrow$		& \fbox{spherical tensor components:  functions of $\EccAno$} 	& $\Leftarrow$ TSH \\
&
to compute time derivatives: use quasi-Keplerian parameterisation
&\\
&
to deal with scalars only: transform to spherical tensor components $I^{lm}$ and $S^{lm}$
&\\
&$\Downarrow$&\\
& \fbox{TD's of $I^{lm}$ and $S^{lm}$ as functions of {\em eccentric anomaly} $\EccAno$}&\\
& $\Downarrow$ &\\
					& decompose and sort appropriately	&	\\
& $\Downarrow$ &\\
\multicolumn{3}{c}{\fbox{TD's of $I^{lm}$ and $S^{lm}$: sum of {\bf even} and {\bf odd} functions of $\EccAno$}} \\
& $\Downarrow$ &\\
&convert (symbolically)&\\
&
even terms: $\sum_{m \ge 0} c_m \cos m \EccAno$,
&\\
&
odd terms: $\sum_{m>0}s_m \sin m \EccAno$,
&\\
&$\Downarrow$&\\
% even terms: cos $\EccAno$ $\Rightarrow$ 
&  
\fbox{TD's  of $I^{lm}$ and $S^{lm}$: sum of trigonometrics of $m \EccAno$ over positive $m$}
&
% odd terms: $\sin \EccAno$
 \\
&$\Downarrow$& \\
&convert (symbolically)& \\
\multicolumn{3}{c}{$\cos m \EccAno$:		~~ $\sum_{j \ge 0} \gamma_j \, \cos j \Mean$,}\\
\multicolumn{3}{c}{$\sin m \EccAno$:		~~ $\sum_{j > 0} \sigma_j \sin j \Mean$,}\\
&($m \in \mathds{N}$) &\\
&$\Downarrow$& \\
\multicolumn{3}{c}{ \fbox{TD's of $I^{lm}$ and $S^{lm}$: sum of trigonometrics of $j \Mean$ over positive $j$}}\\
& $\Downarrow$ & \\
& Fourier transform: straightforwardly performed. &
% & $\Downarrow$ & \\
% $\cos \EccAno$: sum of $\cos j \Mean$ over $j$
% &
% \fbox{components: sum of trigonometrics of $\Mean$}
% &
% $\Leftarrow$
% $\sin \EccAno$: sum of $\sin j \Mean$ over $j$
\end{tabular}
\end{center}
The single steps are being detailled below.

%=======================================================
\section{The binary orbital motion: 2PN accurate Quasi-Keplerian parameterization
%in harmonic coordinates
}
\label{Sec:orbit}
To compute the time derivatives of the multipole moments which we do in harmonic coordinates,
we take the orbital parameterization at 2PN from {\cite{Memmesheimer:Gopakumar:Schafer:2004}},
 
\begin{align}
\label{Eq::orbit_r}
r &= a_r \left ( 1 -e_r\,\cos u \right )\,,\\
\Mean \equiv \MeMo \left ( t - t_0 \right ) &=
u -e_t\,\sin u + 
\epsilon^4
\left[
F_{v-u} \,(v -u)  + F_v  \, \sin v
\right]
 \,,
\label{Eq::Def_KE}
\\
\frac{ 2\,\pi}{ \Phi} \left (\phi - \phi_{0} \right )
&= v + 
\left ( \epsilon^4 {f_{4\phi}}  \right )\,\sin 2v
+
\left ( \epsilon^4 {g_{4\phi}} \right )\, \sin 3v
\,,
% \\
% (\phi-\phi_0)
% &=(1+\periEffective) \, v\,,
\end{align}
% and, to 2PN order, the latter relation can be rewritten as
% \begin{align}
%  (\phi - \phi_0) &= (1+\periEffective) \, v + 
% \left ( \epsilon^4 {f_{4\phi}}  \right )\,\sin 2v
% +
% \left ( \epsilon^4 {g_{4\phi}} \right )\, \sin 3v
% \end{align}
where $ v = 2 \arctan \biggl [ \biggl ( \frac{ 1 + e_{\phi}}{ 1 - e_{\phi}}
\biggr )^{1/2} \, \tan \frac{u}{2} \biggr ]$.
The explicit 2PN accurate expressions for the orbital elements 
and functions of the
generalized quasi-Keplerian parameterization, in harmonic coordinates,
read
\begin{align}
{\it a_r}&= \frac{1}{{(2\En)}} \bigg\{ 1+{\epsilon^2 \frac {{(2\En)}}{4} \left( -7+\eta \right) }
+\epsilon^4 \frac{{{(2\En)}}^{2}}{16} \bigg[1+ {\eta}^{2}
+{\frac{16}{{{(2\En {h}^{2})}}}( {-4+7\,
\eta}}) \bigg]
\bigg\},
\\
{e_{{r}}}^{2}&=
1-{{2\En{h}^{2}}}+
\epsilon^2
{\frac {{(2\En)}}{4\,}} \bigg\{ 24-4
\,\eta+5 \left(-3+ \eta \right) {{(2\En{h}^{2})}} \bigg\}
\no&
+
\epsilon^4
\frac{{{(2\En)}}^{2}}{8}  \bigg\{ 60+148\,\eta+2\,{\eta}^{2} - {{(2\En{h}^{2})}}\left( 
80-45\,\eta+4\,{\eta}^{2} \right)
%\no&&
+\,{\frac {32}{{{(2\En{
h}^{2})}}}(4-7\,\eta)}\bigg\}                       
\,,
\\
%=====================================================
\MeMo &={{(2\En)}}^{3/2} \bigg\{ 1
+
\epsilon^2
{\frac {{{(2\En)}} }{8}}\,(-15+\eta )
+
\epsilon^4
{\frac {{{(2\En)}}^{2}}{128}}\, \bigg[ 555+
30\,\eta
\nonumber \\ &
+11\,{\eta}^{2}+{\frac {192\,}{\sqrt {{{(2\En{h}^{2})}}
}}(-5+2\,\eta)} \bigg] 
\bigg\}, 
\\
%=====================================================
\peri		 &= {
			 \epsilon ^2 \underbrace{\frac{6 \En}{{\OTS}^2}}_{\peri_{\rm 1PN}}
			+\epsilon ^4 \underbrace{\frac{-3 \En^2 \left((9 \eta-22) e_t^2+9 \eta -21\right)}{{\OTS}^4}}_{\peri_{\rm 2PN}}
		} \,, \\
%=====================================================
{e_{{t}}}^{2}&=
1-2\En{h}^{2}+
\epsilon^2
{\frac {{(2\En)}}{4}}\bigg\{ -8+8\,\eta- {{(2\En{h}^{2})}}( -17+7\,\eta)  \bigg\}
\no &
+
\epsilon^4
\frac{{{(2\En)}}^{2}}{8} \bigg\{12+ 72\,\eta+20\,{\eta}^{2}-24 \sqrt {{(2\En h^2)}}
\left( 
-5+2\,\eta \right)
\no&
- {{(2\En{h}^{2})}}( 112-47\,\eta+16\,{\eta}^{2}
)
-{\frac {16}{{{(2\En{h}^{2})}}}(-4+7\,\eta)}
\no&
+{\frac {24}{\sqrt {{(2\En{h}^{2})}}}(-5+2\,\eta)}\bigg\}
\,,
\label{Eq::et_of_j}
\\
F_{v-u} &=
-\frac{3{{(2\En)}}^{2}}{2}\,
\bigg\{{\frac {1}{\sqrt {{(2\En h^2)}}}}(-5+2\,\eta)\bigg\}
\,,
\\                               
F_v &=-\frac{{{(2\En)}}^{2}}{8}\,
\bigg\{
{\frac {\sqrt {1{-2\En}{h}^{2}}}{
\sqrt {{(2\En h^2)}}}\eta\, \left( -15+\eta \right) }
\bigg\},
\\
\Phi&=2\,\pi \, \bigg\{ 1+\epsilon^2{\frac {3}{{h}^{2}}}+
\epsilon^4
\frac{{{(2\En)}}^{2}}{4}
\bigg[ {\frac {3}{{{(2\En{h}^{2})}}}}(-5+2\,\eta)
-{\frac {15}{{{(2\En{h}^{2})}}^{2}}}(-7+
2\,\eta)
\bigg]
 \bigg\}, 
\\
f_{{4 \phi}}&=
\frac{{{(2\En)}}^{2}}{8}
\bigg\{
{\frac {1 {-} {{2 \En{h}^{2}}} }{{{(2\En{h}^{2})}}^{2}}}
 \left( 1+19\,\eta-3\,{\eta}^{2} \right)
\bigg\}
\,,
\\
g_{{4 \phi}}&=
-\frac{{{(2\En)}}^{2}}{32}\bigg\{
{\frac { \left( 1 {-}{{2\En{h}^{2}}} \right) ^{3/2}}{{{(2\En{h}^{2})}}^{2}}}\, \eta\, \left( -1+3\,
\eta \right)                                                                       
\bigg\},
\\
{e_{{\phi}}}^{2}&=
1-2\En{h}^{2}+
\epsilon^2
{\frac {{{(2\En)}}}{4}}\, \bigg\{ 24+ {{(2\En{h}^{2} )}}( -15+\eta ) \bigg\}
\no&
+
\epsilon^4
\frac{{{(2\En)}}^{2}}{16}\,\bigg\{ -40+34\,\eta +18\,{\eta}^{2}-{{(2\En{h}^{2})}} (160 
-31\,
\eta+3\,{\eta}^{2} )
\no&
 -{\frac {1}{{{(2\En{h}^{2})}}}}(-416+91\,\eta+15\,{
\eta}^{2}) \bigg\} 
\,,
\end{align}
where we introduced the abbreviation
\begin{equation}
   \OTS	:=	\sqrt{1-e_t^2}\,.
\end{equation}
There are 2PN accurate relations connecting 
the three eccentricities $e_r, e_t$ and $e_\phi$. These relations
read
\bs
\bea
{ e_t}&=&{ e_r}
\bigg\{
 1+  {\frac {{ (2\En)}}{2{c}^
{2}}}(3\eta-8)+ {\frac {{{ (2\En)}}^{2}}{4{c}^
{4}}}    {\frac {1}{{ (2\En{h}^{2})}}}
 \bigg[
 -16+28\eta
 \no&&
 + (-30+12\eta)\sqrt {{ (2\En h^2)}}
 +(36-19\eta+ 6{\eta}^{2} ) { (2\En{h}^{2})}\bigg]
\bigg\}\,,\\
%==============================
\label{Eq::ephi_er}
{ e_\phi}&=&{ e_r}
\bigg\{
1+  \epsilon^2 {\frac {{ (2\En)}}{2}}  \eta+  \epsilon^4{\frac {{{ (2\En)}}^{2}}{32}
}    {\frac {1}{{ (2\En{h}^{2})}}}
\bigg[ 
160+357\eta-15{\eta}^{2}
\no &&
+(-\eta+ 11{\eta}^{2} ) { (2\En{h}^{2})}
\bigg]
\bigg\}\,.
\eea
\es
%In above equations, the quantity $E$ is defined {\em negative and has unit $c^2$} and, thus, $E\,c^{-2}=- \epsilon^2 \En$ holds
%for the subsequent equations of this paper.
Note that the set ($\En, e_t, \phi_0$) is completely sufficient to describe the initial data of the system in
its plane perpendicular to $\vct{h}$, thus, we
eliminate $h$ for the benefit of $e_t$ in each expression with the help of Equation (\ref{Eq::et_of_j}).
We could also have worked with the QKP in ADM coordinates, but this would have
implied a transformation of the multipole moments. Reference \cite{Gopakumar:Iyer:1997}
provided this transformation for non-spinning compact binaries, coming from a
change in the positions ${\bf x}$ and the velocities ${\bf v}$ at 2PN.
Reference
\cite{Damour:Jaranowski:Schafer:2008:1} {supplied} the associated spin-dependent terms
to 2PN (as we count the spins of order ${\cal O}(\epsilon^0)$).
Working in harmonic coordinates makes life more easy as harmonic coordinates
are a kind of ``natural environment'' for the GW prescription and, there, those
transformations disappear. In the next section, we list the relevant GW moments
and their time derivatives ({thanks to the work of Blanchet, Damour %, Sch\"afer
and Iyer
\cite{Blanchet:Damour:1986, Blanchet:1987, Blanchet:Damour:Iyer:1995}.
Reviews of this subject can be found in \cite{Maggiore:2007} and \cite{Blanchet:2006}}).
%=======================================================
\section{The relevant multipoles and their time derivatives in harmonic coordinates}
\label{Sec::Multipoles}

The far-zone regime GW forms,
regarding only the instantaneous (non-tail) parts,
can be expressed in terms of the mass- and current-type multipole moments
\cite{Thorne:1980, Blanchet:Damour:Iyer:1995},
% {\cite{Thorne:1980}},
\begin{eqnarray}
\label{Eq::HTT_multipoles}
 h^{\rm TT}_{km}(\vct{R}, t) &=& \frac{G}{c^4 R} {\cal P}_{kmij}({\bf N})
 \displaystyle \sum_{l=2}^{\infty} % \sum_{m=-l}^{l} 
\Biggl[
 \left( \frac{1}{c}\right)^{l-2} \left( \frac{4}{l!}\right)
\stackrel{(l)}{\mathcal I}_{ ij {\cal A}_{l-2}} (t-R/c) N_{{\cal A}_{l-2}}
\nonumber \\&&
+ \left( \frac{1}{c}\right)^{l-1} \left( \frac{8l}{(l+1)!}\right)
\epsilon_{pq(i}
 \stackrel{(l)}{\mathcal J}_{j)p {\cal A}_{l-2}} (t-R/c)  \, N_{q{\cal A}_{l-2}}
\Biggr]
\,.
\end{eqnarray}
Here, $A_l$ is a multi-index with $A_l \equiv a_1 ... a_l$, 
indices with round brackets are symmetrized over, viz. for example $A_{\left( ij \right)} \equiv \frac{1}{2} \left( A_{ij} + A_{ji} \right)$, 
and  the following definitions came to use:
\begin{eqnarray}
\vct{R}							&:=& \vct{N} \, R\,, \\
 {\mathcal I}_{{\cal A}_{l-2}}			&:=&	{\mathcal I}_{{a_1}   ... {a_{l-2}}}	\,, \\
  N_{{\cal A}_{l-2}}					&:=&	N_{a_1}  \dots N_{a_{l-2}}								\,,\\
%  P_{ijkl}({\bf n}) 					&:=& 	(\delta_{ik}-N_i N_k)(\delta_{jl}- N_j N_l)-\frac{1}{2}(\delta_{ij} - N_i N_j)(\delta_{kl} - N_k N_l)\,.
P_{ijkl}({\bf N}) 					&:=& 	(\delta_{ik}-N_{i k})(\delta_{jl}-N_{j l})-\frac{1}{2}(\delta_{ij}-N_{i j})(\delta_{kl} - N_{k l})\,.
\end{eqnarray}
$R$ denotes the distance from the observer to the binary.
The quantity $N$ with multiple indices is a tensor product of components $N_i$ 
% \footnote{To avoid confusion with the unit normal vector $\vnunit$ in the orbital plane, we
% wrote $N_{\cal A}$ as a tensor product of $N_{a_l}$ in capital letters.}
of the normal vector
$\vct{N}$ (representing the line of sight from the observer to the center of mass of the binary)
from $N_{a_1}$ to $N_{a_{l-2}}$, and  the superscript $(l)$ denotes
the $l^{\rm th}$ time derivative.
In the current case of 2PN accurate orbital dynamics, we have to evaluate the GW amplitude,
Equation (\ref{Eq::HTT_multipoles}), consistently to 2PN order relative to the leading term, which
kicks in at $\epsilon^4$.
The mass-type multipoles relevant for the above equation read {\cite{Gopakumar:Iyer:1997}}
\begin{eqnarray}
\label{m22}
{\mathcal I}_{ij} &=& \mu\, {\rm STF}_{ij}\biggl \{ x^{ij}
        \biggl [1 +\nonumber \\
       &+& \frac{1}{42\,c^2}\;\left( (29-87\eta)v^2 -
           (30-48\eta)\frac{Gm}{r}\right )\nonumber \\
       &+& \frac{1}{c^4}\left (\frac{1}{504} (253-1835\eta +
           3545\eta^2) v^4\right.\nonumber \\
       &+& \left.\frac{1}{756} (2021-5947\eta -4883\eta^2)
           \frac{Gm}{r}\,v^2\right.\nonumber \\
       &-& \left.\frac{1}{756} (131-907\eta +1273\eta^2)
           \frac{Gm}{r}\,\dot{r}^{2}\right.\nonumber \\
       &-& \left.\frac{1}{252} (355+1906\eta-337\eta^2)
           \frac{G^2m^2}{r^2}\right )\biggr ]\nonumber \\
       &-& x^{i}v^{j}\left[\frac{r\dot{r}}{42\,c^2}
         (24-72\eta)\right.\nonumber\\
       &+& \left.\frac{r\dot{r}}{c^4} \left( \frac{1}{63}
           (26-202\eta +418\eta^2) v^2\right.\right.\nonumber \\
       &+& \left.\left.\frac{1}{378} (1085-4057\eta -1463\eta^2)
           \frac{Gm}{r}\right )\right] \nonumber \\
       &+& v^{ij}\left[ \frac{r^2}{21\,c^2} 
       (11-33\eta)\right.\nonumber\\
       &+& \left.\frac{r^2}{c^4} \left (\frac{1}{126} (41-337\eta
           +733\eta^2)v^2\right.\right.\nonumber \\
       &+&\left.\left.\frac{5}{63} (1-5\eta
           +5\eta^2)\dot{r}^2\right.\right.\nonumber \\
       &+& \left.\left. \frac{1}{189}\, (742-335\eta -985\eta^2)
           \frac{Gm}{r}\right )\right] \biggr \}
\,,
\end{eqnarray}
\begin{eqnarray}
{
{\mathcal I}_{ijk}
}
 &=& -\left(\mu\, \frac{\delta_m}{m} \right)
          {\rm  STF}_{ijk}\Biggl\{ \nonumber \\
       && x^{ijk}\, \Bigl[ 1 + \frac{1}{6\,c^2}
           \Bigl( (5-19\eta )v^2
%  \nonumber \\       &&
  -(5-13\eta ) \frac{Gm}{r} \Bigr) \Bigr] \nonumber \\
       &-& x^{ij}v^k\left [\frac{r\dot{r}}{c^2}
          (1-2\eta )\right ]
       \nonumber \\
       &+& x^iv^{jk}\left [\frac{r^2}{c^2} 
       (1-2\eta )\right ] \Biggr\}
\,,
\\ %============================================
{
{\mathcal I}_{ijkl}
}
	&=& {\mu} \,{\rm STF}_{ijkl}
             \biggl\{ \nonumber \\
        && x^{ijkl}\,\left.\biggl[(1-3\eta ) 
            \right.  \nonumber \\
        &&  
            +\frac{1}{110\,c^2} 
	     \biggl( (103-735\eta
            +1395\eta^2) v^2  \nonumber \\
        &-& (100-610\eta +1050\eta^2)
            \frac{Gm}{r}  \biggr)\biggr] \nonumber\\
        &-& v^ix^{jkl}\left \{\frac{72\,r\,{\dot r}}{55\,c^2} 
	     (1-5\eta +5\eta^2) \right \}
            \nonumber \\
        &+& v^{ij}x^{kl}\left \{\frac{78\,r^2}{55\,c^2}
	(1-5\eta +5\eta^2) \right \}\biggr\}
\,,
\\%===============================================
{\mathcal I}_{ijklm} &=& -\left( \mu \,\frac{\delta_m}{m} \right)\,( 1-2\eta)
             {\rm STF}_{ijklm}\left \{ x^{ijklm}\right \}
\,,  \\
{\mathcal I}_{ijklmn} &=& \mu (1-5\eta +5\eta^2 )
            {\rm STF}_{ijklmn}\left \{  x^{ijklmn}\right \}
\,.
\end{eqnarray}
The current-type moments read
\begin{eqnarray}
{{\mathcal J}_{ij}}
	 &=& -\left(\mu\,\frac{\delta_m}{m} \right)
             {\rm  STF}_{ij} \epsilon_{jab} \Biggl\{
                 \nonumber \\
       && x^{ia}v^b\, \Bigl[1 + \frac{1}{28\,c^2}
           \Bigl( (13-68\eta )v^2 
        \nonumber \\
       &&
        + (54+60\eta )\frac{Gm}{r}\Bigr) \Bigr] \nonumber \\
       &+& v^{ib}x^a \left [\frac{r\,{\dot r}}{28\,c^2} (5-10\eta )
            \right ] \Biggr\}
\,,
\\%========================================
{\mathcal J}_{ijkl} &=& -\left(\mu\,\frac{\delta_m}{m} (1-2\eta )\right )
            {\rm STF}_{ijkl}\left \{
            \epsilon_{lab}\, x^{aijk}v^b \right \}
\,,
 \\
{\mathcal J}_{ijklm} &=& \left ( \mu\, (1-5\eta +5\eta^2)\right ) \,
             {\rm STF}_{ijklm}\left \{
            \epsilon_{mab}\, x^{aijkl}v^b\right \}
\,.
\end{eqnarray}
The notation ${\rm STF}_{ij...}$ denotes the symmetric trace-free part
of the tensor with indices $ij...~$.
The GW amplitude, and from that computed, the far zone angular
momentum and energy transport, has been completed to 3PN in
\cite{Arun:Blanchet:Iyer:Qusailah:2008, Arun:Blanchet:Iyer:Qusailah:2008-1, Arun:Blanchet:Iyer:Sinha:2009}.
From Thorne's paper \cite{Thorne:1980}, see his Equation (4.3), we also extract that
the GW amplitude can equivalently be expressed in terms of tensor
spherical harmonics,
\begin{eqnarray}
 h^{\rm TT}_{jk} &=&  \frac{1}{c^4 R} %{\cal P}_{ jkim}({\bf n})
\displaystyle \sum_{l=2}^{\infty} 
\displaystyle \sum_{m=-l}^{+l}
\Biggl[
\left(  \frac{1}{c}\right)
%  \epsilon
^{l-2}
% \stackrel{(l)}{I} {}^{lm } (t-r/c) T^{E2,lm}_{jk}(\theta,\phi)
{\stackrel{(l )}{I^{lm}}} (t-R/c) T^{E2,lm}_{jk}(\theta,\phi)
\nonumber \\&&
+ 
 \left( \frac{1}{c} \right)
% \epsilon
^{l-1}  
% \stackrel{(l)}{S}{}^{lm} (t-r/c) T^{B2,lm}_{jk}(\theta,\phi)
{\stackrel{(l )}{S^{lm}}} (t-R/c) T^{B2,lm}_{jk}(\theta,\phi)
\Biggr]
\,.
\end{eqnarray}
The components $I^{lm}$ and $S^{lm}$ are projected out of Equation (\ref{Eq::HTT_multipoles}) with the help of the
spherical %tensors
components
 $Y^{lm}_{A_l}$. % (which are symmetric and trace-free in the indices $A_l$),
%which keep the usual name $Y$ of the scalar spherical harmonics, but are suited to expressions with tensorial character.
%which are essential for the basis transformation.
The basis transformation and the explicit representation of the $Y^{lm}_{A_l}$, taking the 
%direction as the conserved
direction of the orbital angular momentum
as the preferred one, $\vAng=h \, \vct{e}_z$, reads
\begin{eqnarray}
 I^{lm}(t)	&=&	\frac{16\pi}{(2\,l+1)!!}		\left[ \frac{(l+1)(l+2)}{2(l-1)l} \right]^{1/2} {\mathcal I}_{A_l}(t) \, Y^{lm*}_{A_l} \,, \\
 S^{lm}(t)	&=&	\frac{-32\pi}{(l+1)(2l+1)!!}	\left[ \frac{(l+1)(l+2)}{2(l-1)l} \right]^{1/2} {\mathcal J}_{A_l}(t) \, Y^{lm*}_{A_l} \,, \\
Y^{lm}_{A_l}	&\stackrel{A}{=}&	(-1)^m (2l-1)!!	\left[ \frac{2l+1}{4\pi (l-m)! \, (l+m)!} \right]^{1/2}
											\left( \delta^1_{\langle i_1} + i \delta^2_{\langle i_1}\right) \dots
											\left( \delta^1_{i_m} + i \delta^2_{i_m} \right)
											\delta^3_{i_{m+1}} \dots
											\delta^3_{i_{l} \rangle } \nonumber \\
			&\stackrel{B}{=}&	(-1)^m \, \left( \frac{2l+1}{4\pi} \frac{(l-m)!}{(l+m)!} \right)^{1/2} \, \sum_{k=0}^{[\frac{1}{2}(l-m)]}
					\frac{(-1)^k (2l-2k-1)!!}{(l-m-2k)!(2k)!!} \times \nonumber \\
				&&	\delta_{(i_1 i_2} \dots \delta_{i_{2k-1} i_{2k}}
					\left(  \delta^1_{i_{2k+1}} + i \delta^2_{i_{2k+1}} \right) \dots
					\left(  \delta^1_{i_{2k+m}} + i \delta^2_{i_{2k+m}} \right) \dots
					\delta^3_{i_{2k+m+1}} \dots \delta^3_{i_l )}\,, \\
Y^{lm*}_{A_l}&=&	(-1)^m Y^{l \, |m| *}_{A_l} \text{~~~for ~} (m<0) \,,
		\label{Eq::Y_lm_negative_m}
\end{eqnarray}
where the notation ``$[x]$'' denotes the integer number which is the largest smaller than or
equal to $x$. From indices comprised by $\langle$ and $\rangle$, the STF parts have to be taken.
The number $l$ tells us what type of moment we have: quadrupole for $l=2$,
octupole for $l=3$, hexadecapole for $l=4$,~\dots~.
From Equation (\ref{Eq::Y_lm_negative_m}) we see that we only have to compute the moments for $m=0 \dots l$.
Equal sign B above, taken from Appendix (A3) of  \cite{Blanchet:Damour:1986},
may be more practical for programming than A.
The relevant pure-spin tensor harmonics, $T^{E2,lm}_{jk}$ and $T^{B2,lm}_{jk}$,
are given in our Appendix \ref{Sec::TensorSphericalHarmonics}.
The reader can find the 1PN outputs in \cite{Junker:Schafer:1992}, which we also used
for our prequel paper.
In case of 2PN accurate conservative dynamics, the GW amplitude, consistently worked out to the same relative
order in $\cinv=\epsilon$ and having moved to our scaling, explicitly reads
\begin{eqnarray}
\label{Eq::HTT_Tens_Spher_Comp}
 h^{TT}_{ij}	&=&
\frac{\epsilon^4}{R}
\Biggl\{
	\sum_{m=-2}^{2} {\stackrel{(2 )}{I^{2m}}} T^{E2,2m}_{ij}
+
\epsilon
	\left[	\sum_{m=-2}^{2} {\stackrel{(2 )}{S^{2m}}} T^{B2,2m}_{ij}+
			\sum_{m=-3}^{3} {\stackrel{(3 )}{I^{3m}}} T^{E2,3m}_{ij}
	\right]
\nonumber \\ && %=====================================
+
\epsilon^2
	\left[	\sum_{m=-3}^{3} {\stackrel{(3 )}{S^{3m}}} T^{B2,3m}_{ij}+
			\sum_{m=-4}^{4} {\stackrel{(4 )}{I^{4m}}} T^{E2,4m}_{ij}
	\right]
\nonumber \\ && %=====================================
+
\epsilon^3
	\left[	\sum_{m=-4}^{4} {\stackrel{(4 )}{S^{4m}}} T^{B2,4m}_{ij}+
			\sum_{m=-5}^{5}  {\stackrel{(5 )}{I^{5m}}} T^{E2,5m}_{ij}
	\right]
\nonumber \\ && %=====================================
+
\epsilon^4
	\left[	\sum_{m=-5}^{5} {\stackrel{(5 )}{S^{5m}}} T^{B2,5m}_{ij}+
			\sum_{m=-6}^{6}  {\stackrel{(6 )}{I^{6m}}} T^{E2,6m}_{ij}
	\right]
\Biggr\}
\,.
\end{eqnarray}
To reach there, we first have to compute the $l^{\rm th}$ time
derivatives of the STF mass and current moments. We obtain them
by means of the accelerations for a compact binary in harmonic
coordinates to 2PN order, 
${\bf a}=
   {\bf a}_{\rm N} 
+ {\bf a}_{\rm 1PN}
+ {\bf a}_{\rm 2PN}$, (also taken from \cite{Gopakumar:Iyer:1997},
written in {\em their units}),
\begin{eqnarray}
%\label{aPN}
%\begin{eqnarray}
{\bf a}_{\rm N} = && - \frac{Gm }{ r^2} {\vnunit} \,, \label{aN}
\\
{\bf a}_{\rm 1PN} = && - \frac{G m }{ c^2\,r^2}
 \biggl\{   \left[
-2(2+\eta)
\frac{Gm }{ r} + (1+3\eta)v^2 - \frac{3}{2} \eta \dot r^2 \right] {\vnunit}
  -2(2-\eta) \dot r {\bf v} \biggr\} \,, \label{a1PN} \\
{\bf a}_{\rm 2PN} = && - \frac{G\,m }{ c^4\, r^2} 
\biggl\{  \biggl[ \frac{3}{4}
(12+29\eta) \frac{G^2\,m^2 }{ r^2} + \eta(3-4\eta)v^4 + \frac{15 }{ 8}
\eta(1-3\eta)
\dot r^4 \nonumber \\
&& - \frac{3 }{ 2} \eta(3-4\eta)v^2 \dot r^2
- \frac{1 }{ 2} \eta(13-4\eta) \frac{G\,m }{ r} v^2 - (2+25\eta+2\eta^2)
\frac{G\,m }{ r} \dot r^2 \biggr]{\vnunit} \nonumber \\
 && - \frac{1 }{ 2}  \left[ \eta(15+4\eta)v^2 -
(4+41\eta+8\eta^2)
\frac{G\,m }{ r} -3\eta(3+2\eta) \dot r^2 \right]\dot r {\bf v} \biggr\} \,.
\label{a2PN}
%\end{eqnarray}
\end{eqnarray}
\noindent
The results are lengthy, easily reproducible and needed only temporarily;
thus, they are not listed here.
We employ the orbital parameterization, Equations~(\ref{Eq::orbit_r}) -- (\ref{Eq::ephi_er}),
to obtain the normal and the velocity vector $\vnunit$ and $\bf v$
in spherical coordinates, symbolically
\begin{eqnarray}
 \vct{r}	&=& \vnunit \, r \,,\label{Eq::vr} \\
{\vnunit}&=&\{\cos (\phi ),\sin (\phi ),0\}\,, \\
\vvelocity:=	\frac{{\rm d} \vct{r}}{{\rm d} t}
	 &=&	\left[\frac{\partial \vct{r}}{\partial u} \right]
		\left[ \frac{\partial u}{\partial \Mean} \right]
		\left[ \frac{\partial \Mean}{\partial t} \right]\label{Eq::vvel} \,.
\end{eqnarray}
%Equation (\ref{Eq::vr})
%using the radial separation of (\ref{Eq::orbit_r}) respectively,
Equation (\ref{Eq::vvel}) is to be computed with the help of the KE (\ref{Eq::Def_KE}).
Again, it is not necessary to provide the
velocity as functions of $u$ and $\phi$ because these terms are easy to be reproduced.
With this input we can compute the spherical tensor components of Equation (\ref{Eq::HTT_Tens_Spher_Comp}).
The next section gives the results, using the 2PN accurate QKP in harmonic coordinates.
%=======================================================
\section{The GW forms: tensor spherical components of $h_{ij}^{TT}$}
\label{Sec:WaveForm}
\noindent
Defining
 \begin{eqnarray}
  A(u)	&:=&	1-e_t\,\cos u \,,
 \end{eqnarray}
we get

{
It will be necessary to decompose these tensor
components in terms of irreducible expressions
to get a time Fourier representation. Those will
be terms %with even and odd parity in $u$, and
%they 
which collect contributions
%of the spherical components
having $\sin u$ on the one hand and
those without $\sin u$ on the other, and they
will be used when we write down the exponential
of the orbital phase in such a way that we can
use results and representations we already know
from the literature or we have to evaluate them from
scratch. This computation is subject of the
next section.
}
%=======================================================
\section{Relevant Kapteyn Series of irreducible components}
\label{Sec:BasicIngredients}
\subsection{Series representation for the inverse KE,
$\sin mu$ and $\cos mu$}

We recall the computation of the 1PN version of this consideration
in paper 1.
There we required only  a Newtonian accurate expression of the sin- and cos-function
of multiples of the eccentric anomaly.
%Because the wave form is
%2PN accurate, we have to provide those terms to the order $\order{\epsilon}{4}$,
%whereas due to the structure of the KE, the 1PN contributions are absorbed in the
%definition of $e_t$ and will not appear.
For further considerations, let us call $u=g(\Mean)$ the solution to the 2PN KE.
As we Taylor expand the argument of the Bessel integral -- which will be done below --
we are in the position to provide $u$ as a series in $\Mean$ up to 2PN.
Therefore we
need the well-known representation of $v$ and $u$ %\colred{\cite{Memmesheimer:Gopakumar:Schafer:2004}}
in
the KE (\ref{Eq::Def_KE}),
% \begin{eqnarray}
%  \Mean	&=&	u-e_t \sin u + \epsilon^4 \, \left( {\cal F}_{v-u} (v-u) +{\cal F}_v \sin v \right)
% \end{eqnarray}
where in the 2PN term, we can insert their Newtonian accurate summation surrogates
(see Equations (5) on p. 553 and (8) on p. 555 in \cite{Watson:1980}. In \cite{Colwell:1993}
there is a misprint in the definition of the $\vKE{n}$ on page 33: the factor $\frac{2}{n}$
should comprise the complete right hand side),
\begin{eqnarray}
\label{Eq::vSumInM}
 v_{(e=e_t)}	&=&\Mean + \sum_{m=1}^{\infty} G_m(e_t) \, \sin m \Mean \,, \\
\label{Def::G_in_J}
 G_m(e)	&=&\frac{2}{m} \left\{ J_m (m e) + \sum_{s=1}^{\infty} \alpha^s \left[ J_{m-s}(me) - J_{m+s}(me)\right] \right\}\,,
\end{eqnarray}
and $\alpha$ is extractable from
\begin{equation}
\label{Def::alpha}
 e = \frac{2\alpha}{1+\alpha^2}\,.
\end{equation}
The series expansions of the functions $(v-u)$ and $\sin v$ at Newtonian accuracy read
\begin{eqnarray}
 (v-u)	&=&	\left( \Mean + \sum_{i=1}^{\infty} \vKE{i}( e_t)\, \sin(i\,\Mean)\right)  - \left( \Mean + \sum_{n=1}^{\infty} \frac{2}{n} J_{n}(n\,e_t ) \sin(n \, \Mean)	\right)	\\
		&=&	\sum_{i=1}^{\infty} \left( \vKE{i}(e_t) - \frac{2}{i} J_{i}(i \,e_t ) \right) \sin(i \, \Mean)\,,\\
\sin v	&=&	\sqrt{1-e_t^2} \, \sum_{n=1}^{\infty} 2 \, J_n^{'}(n\,e_t) \, \sin(n\,\Mean)\,.
\end{eqnarray}
We take above definitions and write in shorthand notation for further calculations,
cf. Equation (\ref{Eq::Def_KE}),
\begin{eqnarray}
\label{Eq::KE_short}
 \Mean &=&  u - e_t\,\sin u  + \epsilon^4\,\sum_{j=1}^{\infty} \alpha_j \, \sin (j \Mean)\,.
\end{eqnarray}
Inserting this in the KE and solving for the Fourier-Bessel coefficients,
 we calculate after Taylor expansion in $\epsilon$
(see Appendix {\ref{Sec::App_KeplerEQ}}),
\begin{eqnarray}
 g(\Mean) - \Mean	&=&	\sum_{n=1}^{\infty} {\cal A}_n \, \sin (n \, \Mean)\,,\\
{\cal A}_n		&=&	\frac{2}{n\,\pi} \int_{0}^{\pi} \cos(n\,\Mean) \, {\rm d} g(\Mean)\nonumber \\
				&=&	\frac{2}{n\,\pi} \int_{0}^{\pi} \cos \left(n\, \left[g_N(\Mean)-e_t\,\sin(g_N(\Mean))
							 + \epsilon^4 \, \sum_{m=1}^{\infty} \alpha_m \sin (\Mean m) \right] \right) \, {\rm d} g(\Mean) \nonumber \\
	&=&					\frac{2}{n\,\pi} \int_{0}^{\pi} \cos \left(n\, \left[g_N(\Mean)-e_t\,\sin(g_N(\Mean)) \right] \right) \, {\rm d} g(\Mean) \nonumber \\
	&&		-\frac{2}{\pi} \int_{0}^{\pi} \sin \left(n\, \left[g_N(\Mean)-e_t\,\sin(g_N(\Mean)) \right] \right) \, \epsilon^4 \, \sum_{m=1}^{\infty} \alpha_m \sin (\Mean m) \, {\rm d} g(\Mean)
\,.
\end{eqnarray}
Defining
\begin{eqnarray}
 \Theta (j,n) := 	\begin{cases}
		0, ~~ j \le n
		\\
		1, ~~ j>n
	\end{cases}
\,,
\end{eqnarray}
the result reads
\begin{eqnarray}
\label{Eq::KE2PN_analytic}
 u	&=& 	\Mean +	2 \left[\sum _{j=1}^{\infty } \frac{\sin (j \Mean) J_j\left(j e_t\right)}{j}\right]
\nonumber \\ && %=====================
-2 \, \epsilon^4 \, {\cal F}_v \sqrt{1-e_t^2}
{\sum_{j=1}^{\infty}}
% \Biggl[
\sin (j \Mean) \times
\nonumber \\&& %=====================
%  old result: some errors occurred
% \Biggl(
% {\sum_{k=1}^{\infty}}
% 	\left[
% 		\rm{J'}_{k}\left(k\, e_t\right) \left(J_{k+j}\left(e_t\, (k+j)\right)-J_{k-j}\left(e_t \,  (k-j)\right)\right)
% 	%,k,1,\infty
% 	\right]
% \nonumber \\&& %=====================
% +{\sum_{m=1}^{j}}\left[
% 				\rm{J'}_{m}\left(m e_t\right) J_{j-m}\left(e_t (j-m)\right)
% %				,m,1,j
% 				\right]
% \Biggr)
% %,j,1,\infty
%  +{J'}_{j}\left(j\,  e_t\right)
% 
% THE NEW ONES:
% ================================================================ 
% ================================================================ 
   \Biggl(
 \left[\sum_{m= 1}^{\infty } J_m\left(e_t m\right){J'}_{j+m}\left(e_t (j+m)\right)\right]
\displaystyle{
-\Biggl[\sum _{m=j+1}^{\infty } J_m\left(e_t m\right) {J'}_{m-j}\left(e_t (m-j)\right)\Biggr]
}
\nonumber \\
&&
+\left[\sum _{m=1}^{j-1}    J_m\left(e_t m\right) {J'}_{j-m}\left(e_t (j-m)\right)\right] \Theta (j,1)
+\text{J'}_j\left(e_t  j \right)\Biggr)
% \Biggr]
%\nonumber \\&& %=====================
%+{\sum_{n=1}^{\infty}}\left[\sin (n M) \rm{J'}_{n}\left(n\,  e_t\right),n,1,\infty\right]
\nonumber \\&& %=====================
+ \epsilon^4 \, {\cal F}_{v-u}
\sum_{j=1}^{\infty}
\sin{j\Mean}
\times \nonumber \\ &&
\Biggl(
\Biggl[\sum _{m=1}^{j-1}
%  \frac{J_m\left(m e_t\right) \left(
%       (m-j) G_{j-m} \left(e_t \right)
% +2            J_{j-m} \left(e_t (j-m) \right)
% \right)}{j-m}
{
-J_m\left(m e_t\right) \left(
G_{j-m}\left(e_t (j-m)\right)
+\frac{2 J_{j-m}\left(e_t (j-m)\right)}{j-m}
\right)
}
\Biggr]
\,\Theta (j,1)
\nonumber \\&& %=====================
+\left[\sum _{m=j+1}^{\infty } J_m\left(m
   e_t\right) \left(G_{m-j}\left(e_t \right)+\frac{2 J_{m-j}\left(e_t
   (m-j)\right)}{j-m}\right)\right]
\nonumber \\&& %=====================
-\left[\sum _{m=1}^{\infty } 
	J_m\left(m  e_t\right) \left(
	G_{j+m}\left(e_t \right)-
\frac{2 J_{j+m}\left(e_t
   (j+m)\right)}
{j+m}
\right)
\right]
\nonumber \\&& %=====================
-G_j\left( e_t\right)+\frac{2 J_j\left(j e_t\right)}{j}
\Biggr)
\,.
%------------------------------------------
\end{eqnarray}
The reader should keep in mind {Appendix C and D} of paper 1
where care is taken of products of infinite series of $\sin j\Mean$
and $\cos j\Mean$
with arbitrary coefficients $\alpha_j$ and $\beta_j$.
This consideration is necessary to collect for terms with
the same positive frequencies in the above expressions and many more.
\\
\noindent
Now we proceed with the trigonometrics of $m\,u$,  ${m \in \mathbb N}$.
We know from the symmetry of $\sin mu$ and $\cos mu$, that only $\sin j\Mean$
or $\cos j\Mean$ can contribute. Thus, we decompose
\begin{eqnarray}
 \sin {mu}	&=&	\sum_{j=1}^{\infty} \SKE{m}{j} \sin {j\Mean} \,, \\
 \cos {mu}	&=&	\sum_{j=0}^{\infty} \CKE{m}{j} \cos {j\Mean} \,.
\end{eqnarray}
The coefficients $\SKE{m}{j}$ and $\CKE{m}{j}$ can be computed using
\begin{eqnarray}
 \CKE{m}{j}	&=&	\frac{2}{\pi} \int_{0}^{\pi} \sin {mu} \, \sin {j\Mean}\, {\rm d} \Mean	\,, \\
 \SKE{m}{j}	&=&	\frac{2}{\pi} \int_{0}^{\pi} \cos{mu} \, \cos {j\Mean}\, {\rm d} \Mean	\,.
\end{eqnarray}
Switching from the integration over d$\Mean$ to d$u$ in the above equations using the 2PN accurate KE
and Taylor expanding everything to $\epsilon^4$, we can perform the integration.
One technical -- but easy to manage -- issue
is to re-convert the arguments of the integrals for a simple application of the
Bessel integral formula,
\begin{equation}
 J_y(x) = \frac{1}{\pi}\int_0^\pi {\rm d}u \cos( y u - x \sin u)\,.
\end{equation}
{Appendix \ref{Appendix::Cos_mu}} provides the calculation.
{The results read }
\begin{align}
\label{Eq::SinMU}
\hspace*{-5cm}
\sin (m\,u) &=
 \sum_{n=1}^{\infty}
% \frac{1}{n} \left\{ J_{n-m}(n\,e_t) + j_{n+m}(n\,e_t)  \right\}
\SKEN{m}{j}
 \, \sin {j\,\Mean}
+ \epsilon^4 \, \sum_{j=1}^{\infty} 
%{^{2}\SKE{m}{j}}
\SKETwoPN{m}{j}
 \, \sin j \Mean \,,\\
\SKEN{m}{j} &:=
\frac{m}{j} \left\{ J_{j-m}(j \,e_t) + J_{j+m}(j \,e_t) \right\}
\,,\\
%^{2}\SKE{m}{j}
\SKETwoPN{m}{j}
% old and mostly wrong:
% &:=&
% \sum_{n=1}^{\infty}
%  \frac{m^2 \kappa _n}{2 j} \, \bigg\{-J_{-j+m-n}\left(e_t (-(j+n))\right)-J_{j+m-n}\left(e_t (j-n)\right)
% \nonumber \\ &
% +J_{-j+m+n}\left(e_t (n-j)\right)+J_{j+m+n}\left(e_t (j+n)\right)\biggr\}
% \,,
% \\
% 
&:=
{
   \frac{1}{2} m 
\sum_{j=1}^{\infty}
\alpha _n \Bigl\{
	-J_{j+m-n}\left(e_t (j-n)\right)
	+J_{j-m+n}\left(e_t (j+n)\right)
}
\nonumber \\ &
{
	-J_{-j+m+n}\left(e_t (n-j)\right)
	+J_{j+m+n}\left(e_t (j+n)\right)
\Bigr\}
}\,,
\\
\label{Eq::CosMU}
\cos (m\,u) &= \,\sum_{j=0}^{\infty}\CKEN{m}{j} \, \cos {j\,\Mean}
+ \epsilon^4 \, \sum_{j=0}^{\infty}
% {^{2}\CKE{m}{j}}
\CKETwoPN{m}{j}
 \, \cos j \Mean \,,\\
\CKEN{m}{j}&:=  \frac{m}{j} \left\{ J_{j-m}(j\,e_t) - J_{j+m}(j \,e_t) \right\} \times \Theta(j,0)  + \delta_{m1}\,\delta_{0j} \left(-\frac{e_t}{2} \right)
\,,
\\
%^{2}\CKE{m}{j}
\CKETwoPN{m}{j}
% &:=&
% also old and mostly wrong:
% \sum _{n=1}^{\infty }
%  -\frac{m^2 \kappa _n}{2 j}
%  \Biggl\{
% J_{-j+m-n}\left(-e_t (j+n)\right)-J_{j+m-n}\left(e_t (j-n)\right)
% \nonumber \\ &
% -J_{-j+m+n}\left(e_t (n-j)\right)+J_{j+m+n}\left(e_t (j+n)\right)
% \Biggr\}
% \nonumber \\ &
% + \delta_{m1}\,\delta_{0j} \left( -\frac{\epsilon ^2 \left(\sqrt{1-e_t^2}-1\right) \left(F_{v-u}+F_v\right)}{e_t} \right)
% \,, \nonumber \\
&:=
{
\frac{1}{2} m \sum_{n=1}^{\infty} \alpha _n \Big\{ J_{-j+m-n}\left(e_t
   (-(j+n))\right)+J_{j+m-n}\left(e_t
   (j-n)\right)
}
\nonumber \\ &
{
-J_{-j+m+n}\left(e_t
   (n-j)\right)-J_{j+m+n}\left(e_t (j+n)\right)\Bigr\} \times \Theta(j,0)
}
\nonumber \\ &
{
+ \delta_{m1}\,\delta_{0j} \left( -\frac{\epsilon ^2 \left(\sqrt{1-e_t^2}-1\right) \left(F_{v-u}+F_v\right)}{e_t} \right)
}
\,.
\end{align}
This prescription is valid in both ADM and harmonic coordinates, where, of course,
the appropriate values always have to be included.

% \begin{flalign*}
% & testLHS1 =& testRHS1  & \hspace{\textwidth} \\
% & testLHS2 =& testRHS2 &
% \end{flalign*}

\subsection{$A(u)^{-n}$ and $A(u)^{-n} \sin u$ as Fourier-Bessel series}

We keep expression (A10) from \cite{Tessmer:Schafer:2010},
\begin{eqnarray}
 A(u)^{-n}			&=&	1 + b^{(n)}_0
%  + \sum_{m=1}^{\infty}  \CKE{m}{0} \, b^{(n)}_m
 + \sum_{j=1}^{\infty} b^{(n)}_{j} \, \cos j u \,, \\
 \frac{\sin u}{A(u)^n}	&=&	\sum_{j=1}^{\infty} S^{(n)}_{j} \sin j u \,.
\end{eqnarray}
In the above two Equations, $b^{(n)}_{j}$ and $S^{(n)}_{j}$ are expansion coefficients for harmonics of $u$. %k,n.jkj
From Equations (\ref{Eq::SinMU}) and (\ref{Eq::CosMU}) we learn that the Fourier-Bessel representation
of the above two series get 2PN corrections,
\footnote{Please recognise the misprint in Equation (39) in paper 1: between $b^{(n)}_0$ and the
round bracket, there should be a ``$+$'' instead of a ``$\cdot$''.}
\hspace{-1cm}
\begin{align}
%\begin{split}
 A(u)^{-n}			
&=
\displaystyle 1 + b^{(n)}_{0} 
+ \sum_{m=1}^{\infty}
%  \gamma^m_0
\CKE{m}{0}
 \, b^{(n)}_m
+ \sum_{j=1}^{\infty} \left( \sum_{k=1}^{\infty} 
% \gamma^{k}_{j}
\CKE{k}{j}
 b^{(n)}_k \right) \cos {j\Mean} 
\nonumber \\
&=
\displaystyle	\sum_{j=0}^{\infty} \AKE{n}{j} \cos {j \Mean}
%\end{split}
 \,, \\
%\begin{split}
\sin u\, A(u)^{-n}		&=\displaystyle	\sum_{j=1}^{\infty} \left( \sum_{k=1}^{\infty}
%  \sigma^{k}_{j}
\SKE{k}{j}
 b^{(n)}_k \right) \sin {j\Mean} 
\nonumber \\			&=	
					\displaystyle	\sum_{j=1}^{\infty} \SAKE{n}{j} \sin {j \Mean}
%\end{split} 
 \,, \\
\AKE{n}{j} 			&=	\delta_{j0} \, \left(1 + b^{(n)}_{0} + \sum_{m=1}^{\infty} \left[  \CKEN{m}{0} + (\epsilon^4) \CKETwoPN{m}{0} \right]\, b^{(n)}_m \right)
							+  \sum_{k=1}^{\infty} \left[ \CKEN{k}{j} + (\epsilon^4) \CKETwoPN{k}{j} \right] b^{(n)}_k  \,,\\
\SAKE{n}{j}			&=  \sum_{k=1}^{\infty} \left[ \SKEN{k}{j} + (\epsilon^4) \SKETwoPN{k}{j} \right] b^{(n)}_k	 \,.
\end{align}

\subsection{Computing one more relevant double series}
We have seen in the 1PN case that we have to deal with products of sine-sine
and sine-cosine series. At 2PN accuracy, we additionally observe products of cosine-cosine
series. The old double series formulas are valid irrespective of the PN order, but
we supply the computation of cosine-cosine series products below. Suppose a term of
the form
\begin{multline}
\label{Eq::Prod_Cos_Cos}
\left(\sum_{k=1}^{\infty} \AKE{n}{k}\,\cos{k\,\Mean}\right)
\left(\sum_{m=1}^{\infty} B_{m} \, \cos{m\,\Mean}\right)
\\
=
\frac{1}{2}
\sum_{k=1}^{\infty} \sum_{m=1}^{\infty}
\AKE{n}{k}\,B_{m}
\left(
\cos\left[(k-m)\Mean\right] + \cos\left[(k+m)\Mean\right]
\right)\,.
\end{multline}
Collecting for contributions with the same the frequency (for $k-m=\pm j$
we obtain $m=k\mp j$ and for part 2, $k+m=j$ we obtain $m=j-k$), the result
reads
\begin{multline}
 \frac{1}{2}
\sum_{k=1}^{\infty} \sum_{m=1}^{\infty}
\left(
\cos\left[(k-m)\Mean\right] + \cos\left[(k+m)\Mean\right]
\right)
\\
=
\frac{1}{2}
\Biggl(
\sum_{k=1}^{\infty} \AKE{n}{k}\,B_k
+
\sum_{j=1}^{\infty}
\cos{j\Mean}\times
%\nonumber \\ 
\left[
   \sum_{k=j+1}^{\infty}	\AKE{n}{k} \, B_{k-j}
+\sum_{k=1}^{\infty}		\AKE{n}{k} \, B_{k+j}
+\sum_{k=1}^{j-1}		\AKE{n}{k} \, B_{j-k} \, \Theta(j,1)
\right]
\Biggr)
\\
=:
\sum_{j=0}^{\infty}
\ProdCC{j} \,
\cos{j\Mean}\,.
\end{multline}

\subsection{Decomposition of exp$\left\{ - i m \phi		\right\}$}
What we have done at 1PN accuracy has to be extended to 2PN, especially at the orbital dynamics.
It is helpful to find a special decomposition of $e^{-i m \phi}$ in such a way that the mode
decomposition of any 1PN function of $u$ (to be performed exactly) is not required at this point of calculation. We will combine the
terms in such a way that we can use results known from the previous sections,
\begin{align}
 (\phi - \phi_0) &= (1+\periEffective) \, v + 
\epsilon^4 \, 
\left(
{f_{4\phi}} \, \sin 2v
+
{g_{4\phi}}\, \sin 3v
 \right)
\,,
\\
v=v(u)	
&\stackrel{1}{=}
	2\, 
	\arctan \left\{ \sqrt{\frac{1+e_\phi}{1-e_\phi}}
			\tan \left[ \frac{u}{2} \right]
		\right\} \nonumber \\
&\stackrel{2}{=}
	v_{\rm N} + \epsilon^2 \, v_{\rm 1PN} + \epsilon^4\,v_{\rm 2PN} \nonumber \\
&\stackrel{3}{=}:	(\Mean+\tilde{v}_{\rm N})+ \epsilon^2 \, v_{\rm 1PN} + \epsilon^4\,v_{\rm 2PN} \,, \\
% \begin{split}
 e^{-i\,m\, (\phi - \phi_0)}	
&\stackrel{4}{=}	e^{	-i\,m\,  
		 \left[ (1+\periEffective) v
		 + \left ( \epsilon^4 {f_{4\phi}}  \right )\,\sin 2v
		 + \left ( \epsilon^4 {g_{4\phi}} \right )\, \sin 3v
		\right]	
		}
\nonumber \\ %=====================
&\stackrel{5}{=}	e^{-i\,m\, v} \,
	e^{-i\,m (\epsilon^2 \peri_{\rm 1PN} + \epsilon^4 \peri_{\rm 2PN}) \,
		\left[ v_{\rm N} + \epsilon^2 v_{\rm 1PN} + \epsilon^4 v_{\rm 2PN}\right]
	 }
	e^{-i\,m \, \epsilon^4
		 \left[
		   {f_{4\phi}} \,\sin 2v
		 + {g_{4\phi}} \,\sin 3v
		\right]
}
\nonumber \\ %=====================
&\stackrel{6}{=}	e^{-i\,m\, v} \,
	e^{-i\,m {
		v_{\rm N}} \epsilon ^2
		\left({\peri_{\rm 1PN}}+{\peri_{\rm 2PN}} \epsilon ^2\right)} \,
% 	e^{-im {\peri_{\rm 1PN}} {v_{\rm 1PN}} \epsilon ^4}
	e^{-i\,m \, \epsilon^4
		 \left[
		  \peri_{\rm 1PN} {v_{\rm 1PN}} 
		+ {f_{4\phi}} \,\sin 2v
		+ {g_{4\phi}} \,\sin 3v
		\right]
}
 \nonumber \\ %=====================
&\stackrel{7}{=}
	%e^{-i\,m\, v} \,
	e^{-i\,m\, (\epsilon^2 \peri_{\rm 1PN} + \epsilon^4 \peri_{\rm 2PN})\,  \Mean} \times
 \nonumber \\ %=====================
& ~\	\underbrace{ e^{-i\,m\, v}}_{{\rm Part}_{\rm A}(m)} \, 
	\underbrace{\left[
	1- i m \peri_{\rm 1PN} {\tilde{v}_N} \epsilon ^2
		- \frac{1}{2} m {\tilde{v}_N} \epsilon ^4 \left(m {\peri_{\rm 1PN}}^2 {\tilde{v}_N}+2 i \peri_{\rm 2PN} \right)\right]
									}_{{\rm Part}_{\rm B}(m)} \times
\nonumber \\ %=====================
 &~
%  \underbrace{\left[ 1-i m \peri_{\rm 1PN} v_{\rm 1PN} \epsilon ^4 \right]}
% _{=:{\rm Part}_{\rm C}(m)}\,.
\underbrace{
\left[
1-i m \epsilon ^4 (
 {\peri_{\rm 1PN}} {v_{\rm 1PN}}
+{f_{4\phi}} \sin (2 {v_{\rm N}})
+{g_{4\phi}} \sin (3 {v_{\rm N}})
)
\right]
}
_{{\rm Part}_{\rm C}(m)}\,.
% \end{split}
\end{align}
% Mark: new explanation Partb
{
In step 3, $v_{\rm N}$ decomposes into a purely secular part, namely $\Mean$,
and a purely oscillatory one with zero average over the interval $\Mean=[0,2\pi]$, viz. $\tilde{v}_{\rm N}$
[see Equation (\ref{Eq::vSumInM})].
Therefore, in step 7, the middle term in edgy brackets (``Part$_{\rm B}$'') can be written in terms of 
single and double summations of terms in the right hand side of Equation (\ref{Eq::vSumInM}), almost without
computational cost.
% This is the case because we need $v$ only at Newtonian order (which is known as a sum in $\sin j\Mean$).
% The product of the first and the third coefficient can be written as functions of $A(u)$ and $\sin u$, but as the result
% is quite lengthy and also required only temporarily, we will not list it here.
% Recall that at 1PN accurate dynamics, the last part is missing.
Part$_{\rm A}$ and Part$_{\rm C}$ will contribute
$A(u)^{-n}$ and also $A(u)^{-n}\,\sin u$ terms that will
multiply with the series of Part$_{\rm B}$.
These contributions are lengthy and we again skip explicit expressions.
In principle, other decompositions are valid as well,
but we find it convenient to choose the above one
because of its structural clearness.
% to see clearly where higher post Newtonian orders and
% contributions with and without average enter.
Part$_{\rm B}$ is essential and we give it explicitly,
\footnote{{In Equation (C5) of paper 1, the very last term should get a factor $\Theta(j,1)$ to
make it consistent with our notation and
for convenience of the reader.}}
%We recall the definition for the coefficients $\vKE{j}$ in the Fourier expansion of $v$, section III in paper 1 and obtain
}
%We recall the definition for the coefficients $\vKE{j}$ in the Fourier expansion of $v$, section III in paper 1 and obtain
\begin{eqnarray}
%\begin{split}
{\rm Part}_{\rm B} (m) &:=&
\left[1	- i m \peri_{\rm 1PN} {\tilde{v}_{\rm N}} \epsilon ^2
											- \frac{1}{2} m {\tilde{v}_{\rm N}} \epsilon ^4 \left(m {\peri_{\rm 1PN}}^2 {\tilde{v}_{\rm N}}+2 i \peri_{\rm 2PN} \right)\right]
\nonumber  \\
%==========================================================================
 &=&
1
 -i m \epsilon^2\, {\peri_{\rm 1PN}} \left[\sum
   _{j=1}^{\infty } \sin (j \Mean) \, G_j\left( e_t\right)\right]
\nonumber  \\
 &&
- \epsilon^4 \, \frac{1}{4} m 
\Biggl\{_1
\left[\sum _{j=1}^{\infty } m
  \peri_{\rm 1PN}^2 G_j\left( e_t\right){}^2\right]
\nonumber \\ &&
+ \sum_{k=1}^{\infty }
 \Biggl[_2 m\peri_{\rm 1PN}^2 \cos (k \Mean)
\times
% \nonumber \\ && 
\Biggl(
\left[\sum_{j=k+1}^{\infty } G_j\left( e_t\right)
   G_{j-k}\left(e_t \right)\right]
 \nonumber \\ && \qquad %=================
\qquad\qquad\qquad\qquad\quad
-\left[\sum
   _{j=1}^{k-1} G_j\left( e_t\right)
   G_{k-j}\left(e_t \right)\right] {\times \Theta(k,1)}
 \nonumber \\ && \qquad %=================
\qquad\qquad\qquad\qquad\quad
+\left[\sum_{j=1}^{{\infty}} G_j\left( e_t\right)
   G_{j+k}\left(e_t \right)\right]\Biggr)
\nonumber \\ && \qquad %=================
\quad +4 i {\peri_{\rm 2PN}} \sin
   (k \Mean) G_{k}\left( e_t\right)
\Biggr]_2
\Biggr\}_1
%\end{split}
\,.
\end{eqnarray}
{For a clear understanding, we sometimes have added auxiliary indices to the brackets.
This simply helps to see how deep the bracket in the current expression is.}
To 1PN order we recognize what we computed in paper 1, Section IV.
We will face products of
% what we simply call
Part$_{\rm B}$  with powers of $A(u)$ .
%  This product is given below.
They can be put into the form
\begin{eqnarray}
%\begin{split}
\label{Eq::PartB_An}
\frac{{\rm Part}_{\rm B}(q)}{A(u)^n}
&=&
\sum _{j=1}^{\infty } \frac{1}{2} i q \epsilon ^2   \left(\peri_{\rm 1PN}+\peri_{\rm 2PN} \epsilon ^2\right) \, \sin (j \Mean) 
\times  \nonumber \\ &&
\quad 
\Biggl\{\sum_{m=1}^{j-1}	\Biggl[-\AKE{n}{m} \Theta (j,1) G_{j-m}\left(e_t \right)\Biggr]
+\sum_{m=j+1}^{\infty }	\Biggl[\AKE{n}{m} G_{m-j}\left(e_t \right)\Biggr]
\nonumber  \\& &
+\left[\sum_{m=1}^{\infty } -\AKE{n}{m} G_{j+m}\left(e_t \right)\right]
-2 \AKE{n}0 G_j\left(e_t\right)
\Biggr\}
+
\nonumber  \\
\hline%=======================================================================
&&
-\frac{1}{8}
\sum _{ j=1}^{\infty }
 \cos   ( j \Mean)
 \Biggl\{_1
\peri_{\rm 1PN}^2 q^2 \epsilon ^4
   \Biggl[_1
\nonumber  \\ &&
    \sum _{k= j+1}^{\infty }
   \Biggl[
    \AKE{n}{k- j}
   \Biggl(- \sum _{m=1}^{k-1} \left[ G_m\left( e_t \right)
   G_{k-m}\left( e_t \right)\right]
%\nonumber  \\ &&
  + \sum_{n=k+1}^{\infty } \left[G_n\left(e_t\right) G_{n-k}\left(e_t\right)\right]
\nonumber  \\ &&
+ \sum _{n=1}^k \left[ G_n\left( e_t\right) G_{k+n}\left(e_t\right)\right]
\Biggr)\Biggr]
\nonumber  \\ && % --------------------------------------------------------
+ \sum _{k=1}^{\infty}
   \Biggl[
	\AKE{n}{ j+k }  \Biggl(
	- \sum _{m=1}^{k-1} \left[ G_m\left( e_t\right) G_{k-m}\left(e_t \right)\right]
	+ \sum_{n=k+1}^{\infty }\left[ G_n\left( e_t\right) G_{n-k}\left(e_t \right)\right]
\nonumber  \\ && % --------------------------------------------------------
	+ \sum _{n=1}^k \left[ G_n\left( e_t\right) G_{k+n}\left(e_t \right)\right]
	\Biggr)\Biggr]
\nonumber  \\ &&% --------------------------------------------------------
+\Theta (j,1) 
\sum _{s=1}^{ {j}-1} \Biggl[
   \AKE{n}{j-s} \Biggl(- \sum _{m=1}^{s-1} \left[ G_m\left( e_t\right) G_{s-m}\left(e_t \right)\right]
%\nonumber  \\ &&% --------------------------------------------------------
+ \sum _{n=s+1}^{\infty } \left[ G_n\left( e_t\right) G_{n-s}\left(e_t \right) \right]
\nonumber  \\ &&% --------------------------------------------------------
+ \sum _{n=1}^s \left[ G_n\left( e_t\right) G_{n+s}\left(e_t \right)\right]\Biggr)\Biggr]
 \nonumber  \\ &&% --------------------------------------------------------
+2 {\AKE{n}{0} }
   \Biggl(- \sum _{m=1}^{ j-1} \left[ G_m\left( e_t\right) G_{ j-m}\left(e_t \right)\right]
  + \sum_{n= j+1}^{\infty } \left[ G_n\left( e_t\right) G_{n- j}\left(e_t \right)\right]
\nonumber  \\ &&% --------------------------------------------------------
+ \left[\sum _{n=1}^{ j}
   G_n\left( e_t\right) G_{ j+n}\left(e_t \right)\right]
   \Biggr)
\Biggr]_1
% \nonumber  \\ && % --------------------------------------------------------
+2 \AKE{n}{j} \left(\peri_{\rm 1PN}^2
   q^2 \epsilon ^4  \left[\sum _{k=1}^{\infty } G_k\left( e_t\right)^2\right]-4\right)
\Biggr\}_1
\nonumber  \\
\hline
 &&
+\Biggl\{\sum _{j=1}^{\infty } \Biggl[ \frac{1}{8}
   {\peri_{\rm 1PN}}^2 q^2 \epsilon ^4 
\Biggl(\AKE{n}{j}
   \Biggl(-\sum _{k=j+1}^{\infty } \left[ G_k\left( e_t\right) G_{k-j}\left(e_t  \right)\right]
\nonumber  \\&&
+\sum _{k=1}^{j-1} \left[
   G_k\left( e_t\right) G_{j-k}\left(e_t \right)\right]
-\sum _{k=1}^{j}\left[
   G_k\left( e_t\right) G_{j+k}\left(e_t \right)\right]\Biggr)
\nonumber  \\ &&
-2 {\AKE{n}0} G_{j}\left( e_t\right)^2
\Biggr)\Biggr]+{\AKE{n}0}
\Biggr\}
\,.
%\end{split}
\end{eqnarray}
% As usual, we also give the version with $\sin u$,
The part including $\sin u$ reads
\begin{eqnarray}
% \begin{split}
\label{Eq::PartB_An_SinU}
\frac{{\rm Part}_{\rm B}(q)}{A(u)^n}\,\sin u
&&=
\sum _{j=1}^{\infty } \frac{1}{2} \sin (j \Mean) \times
\nonumber \\ &&
   \Biggl\{
2  \SAKE{n}{j}
+	\sum _{m=1}^{j-1}
	 \Biggl[-\frac{1}{4} \peri_{\rm 1PN}^2 q^2 \epsilon ^4  \Theta(j,1) \SAKE{n}{j-m} \Biggl(
\sum _{k=m+1}^{\infty } \Bigl[ G_k\left( e_t\right) G_{k-m}\left(e_t \right)\Bigr]
\nonumber  \\ && %==========================================
+\sum _{k=1}^{m-1}
\Bigl[   -G_k\left( e_t\right) G_{m-k}\left(e_t \right)\Bigr]
+\sum_{k=1}^m \Bigl[ G_k\left( e_t\right) G_{k+m}\left(e_t \right)\Bigr]\Biggr)\Biggr]
\nonumber  \\ && %==========================================
+\sum _{m=j+1}^{\infty }
   \Biggl[
    \frac{1}{4} \peri_{\rm 1PN}^2 q^2 \epsilon ^4 \SAKE{n}{m-j}
	 \Biggl(
	\sum _{k=m+1}^{\infty }	\left[ G_k\left( e_t\right) G_{k-m}\left(e_t \right)\right]
\nonumber  \\ && %==========================================
+	\sum _{k=1}^{m-1}		\Bigl[ -G_k\left( e_t\right) G_{m-k}\left(e_t \right)\Bigr]
+	\sum _{k=1}^m \Bigl[ G_k\left( e_t\right) G_{k+m} \left(e_t \right)\Bigr]
	\Biggr)
   \Biggr]
\nonumber  \\ && %==========================================
+	\sum_{m=1}^{\infty }\Biggl[ -\frac{1}{4} \peri_{\rm 1PN}^2 q^2 \epsilon ^4 \SAKE{n}{j+m}
   \Biggl(\left[\sum _{k=m+1}^{\infty } G_k\left( e_t\right)
   G_{k-m}\left(e_t \right)\right]
\nonumber  \\ && %==========================================
+\left[\sum _{k=1}^{m-1}-	G_k\left( e_t\right) G_{m-k}	\left(e_t \right)\right]
+\left[\sum _{k=1}^{m   }	G_k\left( e_t\right) G_{k+m}	\left(e_t \right)\right]\Biggr)\Biggr]
\Biggr\}
%========================================
\nonumber  \\ 
&& +\left[\sum _{j=1}^{\infty } G_j\left( e_t\right)^2\right] \left[\sum _{j=1}^{\infty } -\frac{1}{4}
   \peri_{\rm 1PN}^2 q^2 \epsilon ^4 \SAKE{n}{j} \sin (j M)\right]
\nonumber  \\
\hline %========================================
&&
-\frac{1}{2} i q \epsilon ^2 \left(\peri_{\rm 1PN}+\peri_{\rm 2PN} \epsilon ^2\right)
   \sum _{j=1}^{\infty } \cos (j \Mean) \times  \nonumber  \\ &&
\Biggl\{
   \Biggl(
		\sum _{m=j+1}^{\infty }	\Bigl[	\SAKE{n}{m-j} G_m\left( e_t\right)\Bigr]
	+	\sum _{m=1}^{\infty }		\Bigl[	\SAKE{n}{j+m} G_m\left( e_t\right)\Bigr]
\Biggr)
\nonumber  \\ &&
-\Theta (j,1)	\sum _{m=1}^{j-1}		\Bigl[ 	\SAKE{n}{j-m}  G_m\left( e_t\right)\Bigr]
\Biggr\}
\nonumber  \\
%=================================================
\hline
&&
-\frac{1}{2} i q \epsilon ^2 \left(\peri_{\rm 1PN}+\peri_{\rm 2PN} \epsilon ^2\right)
  \left[\sum _{j=1}^{\infty } \SAKE{n}{j} G_j\left( e_t\right)\right]
\,.
%\end{split}
\end{eqnarray}
This will be shortened by writing
\begin{eqnarray}
\frac{{\rm Part}_{\rm B}(q)}{A(u)^n}
&=&
\left\{
% \funcAKEc{n,q}_0
    \sum_{j=0}^{\infty} 	\funcAKEc{n,q}_ j  \cos j \Mean
+ \sum_{j=1}^{\infty} 	\funcAKEs{n,q}_ j  \sin j \Mean
\right\}
%e^{-iq ( \epsilon^2 \peri_{\rm 1PN} + \epsilon^4 \peri_{\rm 2PN} ) \Mean}
\,,	
\label{Eq::Separate_Ainvnq_M}
\\
%=================================
\sin u \, \frac{{\rm Part}_{\rm B}(q)}{A(u)^n}
&=&
\left\{
%\funcSinAKEc{n,q}_0
    \sum_{j=0}^{\infty} 	\funcSinAKEc{n,q}_ j  \cos j \Mean
+ \sum_{j=1}^{\infty} 	\funcSinAKEs{n,q}_ j  \sin j \Mean
\right\}
%e^{-iq ( \epsilon ^2 \peri_{\rm 1PN} +  \epsilon^4 \peri_{\rm 2PN} ) \Mean}
 \,,
% \label{Eq::Separate_SinAinvnq_M}
% \\
% %=================================\colred{\AKE{n}{0}} 
% \funcAKEs{n,q}_{j>0}			&:=&	(-i q \peri) \left[ \colred{\AKE{n}{0}}  \vKE{j} +  \ProdCS{j} \right]	\,,	\\
% % (-i q \peri) \left[ (1+b^{(n)}_{0}) \vKE{j} +  \ProdCS{j} \right]	\,,	\\
% \funcAKEc{n,q}_{0}			&:=&	 \left[(1+b^{(n)}_{0})		 					  \right]	\,,	\\
% \funcAKEc{n,q}_{j>0}			&:=&	\AKE{n}{j}\,, \\%  b^{(n)}_{j}										\,,	\\
% %\colblu{
% %\funcAKEc{n,q}_{j	}}			&:=&	\colblu{\delta_{j,0}+b^{(n)}_{j}}							\,,	\\
% \funcSinAKEc{n,q}_{j>0}		&:=&	(-i q \peri) \ProdSS{j}	\,,	\\
% \funcSinAKEs{n,q}_{j>0}		&:=&	\SAKE{n}{j}	\,.
\end{eqnarray}
with $\funcAKEc{n,q}$, $\funcAKEs{n,q}$, $\funcSinAKEc{n,q}$ and $\funcSinAKEs{n,q}$ to be easily extracted from
Equations (\ref{Eq::PartB_An}) and (\ref{Eq::PartB_An_SinU}) and as well remembering that
$\AKE{n}{j}$ and $\SAKE{n}{j}$ have 2PN terms. Equations (\ref{Eq::PartB_An}) and
(\ref{Eq::PartB_An_SinU}) appear to be relatively complicated, but they are simply
expanded applications of the product formulas for $\sin$ and $\cos$ series,
{Equations (C1) and (D1) of paper 1, and our current Equation (\ref{Eq::Prod_Cos_Cos})}.
As in paper 1, we can now decompose the wave form into the above irreducible components,
from whose we have extracted the time Fourier series representation.
% , which read
% $\funcAKEs{n,q}_j$,
% $\funcAKEc{n,q}_j$,
% $\funcSinAKEs{n,q}_j$, and
% $\funcSinAKEc{n,q}_j$.
To simplify matters, we introduce
\begin{eqnarray}
\frac{{\rm Part}_{\rm B}(q)}{A(u)^n} &=:&\funcAnq{n}{q}%	\, e^{-iq\peri\Mean}
 \,,
\label{Def::FuncNQ}	\\
\sin u \, \frac{{\rm Part}_{\rm B}(q)}{A(u)^n}&=:&\funcSinAnq{n}{q} %	\, e^{-iq\peri\Mean}
 \,.
\label{Def::FuncSinNQ}
\end{eqnarray}
%=======================================================
\subsection{Multipole moment decomposition: a brief posting of the results}
\label{Sec::Results}

Using the results of the previous section, we decompose the 
multipole coefficients $I$ and $S$ as
\begin{eqnarray}
 \stackrel{(a)}{{I}^{am}} &=& {\cal C}_{ I;am} \, e^{-m i \peri \Mean} \, e^{-m i \phi_0}\times
\left\{
	\sum_{k} {^{a-2} }\alpha_{[k m]}	\funcAnq{k}{m}
+	\sum_{k'} {^{a-2} }\salpha_{[k m]}	\funcSinAnq{k'}{m}
\right\}
\,,
\\
 \stackrel{(b)}{{S}^{bm}} &=& {\cal C}_{ S;bm} \, e^{-m i \peri \Mean} \, e^{-m i \phi_0}\times
\left\{
	\sum_{k} {^{b-1} }\beta_{[k m]}	\funcAnq{k}{m}
+	\sum_{k'} {^{b-1} }\sbeta_{[k m]}	\funcSinAnq{k'}{m}
\right\}
\,,
\end{eqnarray}
symbolically,
where ${\cal C}_{ I;am}$ and ${\cal C}_{ S;am}$ are some pre-factors and $\alpha, \salpha$,
$\beta$ and $\sbeta$ are coefficients to be determined, $k$ and $k'$ are some summation dummy
indices with boundaries depending on $a$ and $m$ and depending on the type ($I$ or $S$),
and ${{a-2}}$ and ${{b-1}}$ are labels for $\alpha, \salpha$ and for $\beta, \sbeta$,
counting the order of $\cinv$.
The associated components of $I$ and $S$, including ${\rm Part}_{\rm A}$ and ${\rm Part}_{\rm C}$, are given in Appendix~\ref{Appendix::Results}.
Some ``pre -- Fourier'' domain reads

\vspace{0.3cm}
\hspace{-0.4cm}
\fbox{
\begin{minipage}[!hr]{0.95\textwidth}
\begin{eqnarray}
\label{Eq::Fou_I_Mixed}
  \stackrel{(a)}{{I}^{am}} &=&
{\cal C}_{ I;am} \, e^{-m i \peri \Mean} \, e^{-m i \phi_0}\times
% \nonumber \\ &&
\Biggl\{
	\sum_{j} 	\sin{j \Mean} 
\, \stackrel{(a)}{{I}^{am}_{Sj}}%_{j}
% \left( \sum_{k} \alpha_{[k m]} \funcAKEs{n,q}_j
%  +\sum_{k'} \salpha_{[k' m]} \funcSinAKEs{k',q}_j \right)
% \nonumber \\ &&
+	\sum_{j} 	\cos{j \Mean}
\, \stackrel{(a)}{{I}^{am}_{Cj}}%_{j}
%   \left( \sum_{k} \alpha_{[k m]} \funcAKEc{n,q}_j
%  +\sum_{k'} \salpha_{[k' m]} \funcSinAKEc{k',q}_j \right)
\Biggr\}
\,,
\\
%===============
\label{Eq::Fou_S_Mixed}
  \stackrel{(b)}{{S}^{bm}} &=&
{\cal C}_{ S;bm} \, e^{-m i \peri  \Mean} \, e^{-m i \phi_0}\times
% \nonumber \\ &&
\Biggl\{
	\sum_{j} 	\sin{j \Mean}
\, \stackrel{(b)}{{S}^{bm}_{Sj}}%_{j}
%  \left( \sum_{k} \beta_{[k m]} \funcAKEs{n,q}_j
%  +\sum_{k'} \sbeta_{[k' m]} \funcSinAKEs{k',q}_j \right)
% \nonumber \\ &&
+	\sum_{j} 	\cos{j \Mean}
\, \stackrel{(b)}{{S}^{bm}_{Cj}}%_{j}
%   \left( \sum_{k} \beta_{[k m]} \funcAKEc{n,q}_j
%  +\sum_{k'} \sbeta_{[k' m]} \funcSinAKEc{k',q}_j \right)
\Biggr\}
\,,
\end{eqnarray}
with
\begin{eqnarray}
\stackrel{(a)}{{I}^{am}_{Sj}}%_{j}
% \stackrel{(a)}{I}{}^{am}_S
&:=&
\left(		\sum_{k}{} 	{^{a-2} }\alpha_{[k m]} \funcAKEs{k,m}_j
 +		\sum_{k'}{}	{^{a-2} }\salpha_{[k' m]} \funcSinAKEs{k',m}_j \right)
\,, \\
\stackrel{(a)}{{I}^{am}_{Cj}}%_{j}
&:=&
  \left(	\sum_{k}^{}	{^{a-2} } \alpha_{[k m]} \funcAKEc{k,m}_j
 +		\sum_{k'}^{}	{^{a-2} } \salpha_{[k' m]} \funcSinAKEc{k',m}_j \right)
\,, \\
\stackrel{(b)}{{S}^{b m}_{Sj}}%_{j}
&:=&
 \left(	\sum_{k}{}	{^{b-1} }\beta_{[k m]} \funcAKEs{k,m}_j
 +		\sum_{k'}{}	{^{b-1} }\sbeta_{[k' m]} \funcSinAKEs{k',m}_j \right)
\,, \\
\stackrel{(b)}{{S}^{b m}_{Cj}}%_{j} 
&:=&
  \left(	\sum_{k}		{^{b-1} }\beta_{[k m]} \funcAKEc{k,m}_j
 +		\sum_{k'}	{^{b-1} }\sbeta_{[k' m]} \funcSinAKEc{k',m}_j \right)
\,,\\
{\rm for}&& a \in [2,6] \,, b \in [2,5] \,, \nonumber
\end{eqnarray}
\end{minipage}
}
 \vspace{0.3cm}

\noindent
and for extracting the pure Fourier domain representation with delta distributions
-- (and not the one mixed in exponential and trigonometric representation
as in Equations (\ref{Eq::Fou_I_Mixed}) and (\ref{Eq::Fou_S_Mixed})) --
we take the Fourier transformation of the $\sin j \Mean$ and $\cos j\Mean$ terms,
\begin{eqnarray}
\frac{1}{\sqrt{2 \pi}}
 \int_{-\infty}^{+\infty} % e^{-i\magnetic\phi_0}
e^{-i \magnetic \peri \Mean} \sin {j\Mean} \, e^{i \omega t} {\rm d}t&=&
i \sqrt{\frac{\pi }{2}} %e^{-i m {\phi_0}}
	\delta (j \MeMo+\peri \magnetic \MeMo-\omega)
-i \sqrt{\frac{\pi }{2}} %e^{-i m {\phi_0}}
	\delta (j \MeMo - \peri \magnetic \MeMo+\omega)\,, \\
\frac{1}{\sqrt{2 \pi}}
 \int_{-\infty}^{+\infty} %e^{-i\magnetic\phi_0}
 e^{-i \magnetic \peri \Mean} \cos {j\Mean} \, e^{i \omega t} {\rm d}t&=&
\sqrt{\frac{\pi }{2}}	\delta (j \MeMo+\peri \magnetic \MeMo -\omega)
+\sqrt{\frac{\pi }{2}}	\delta (j \MeMo- \peri \magnetic \MeMo+\omega)\,.
\end{eqnarray}
This  is valid only in the purely conservative orbital dynamics case.

\subsection{The effect of radiation reaction}
Up to now, we have used only the conservative parameterization
of Section \ref{Sec:orbit} hitherto. When we include the slow-in-time
variation of the orbital elements due to radiation reaction, the Equations
(\ref{Eq::Fou_I_Mixed}) and (\ref{Eq::Fou_S_Mixed}) are no longer
valid, because the radial frequency $\MeMo$ is no longer a constant.
In paper 1, we gave a detailled overview over what to do when eccentric GW
signals with chirp have to be Fourier transformed using the
algorithm of the steepest decent or equivalently called
``the method of the stationary phase''. The only difference to
be recognized is that we may include 2PN accurate formulas
for $\dot {e}_t$ and $\dot{\MeMo}$ and the QKP, but the analytic integration of those EOM
will be skipped because of the extensive space filling
of the results as we have faced already at 1PN. 
A future publication may extend the calculation of {Section VI} of paper 1
with the help of Appell's integral formula (see e.g. \cite{Olver:Lozier:Boisvert:Clark:2010}),
the {\em orbital averaged} 2PN accurate EOM
\begin{eqnarray}
 \left\langle \frac{{\rm d} \MeMo  }{{\rm d} t} \right\rangle_{\rm orbit} &=& {\mathfrak N}_{\rm 2PN}(\MeMo, e_t)\,,\\
 \left\langle \frac{{\rm d} e_t}{{\rm d} t} \right\rangle_{\rm orbit} &=& {\mathfrak E}_{\rm 2PN}(\MeMo, e_t)\,,
\end{eqnarray}
to be taken from the expressions (4.32) and (4.34) of reference \cite{Gopakumar:Iyer:1997}
and converted appropriately to $\MeMo$ and $e_t$, and the
perturbation routines for inverting the formulas for the time of steepest decent we are familiar with.

The interested reader may find the stationary phase approximation (SPA) results for circular orbits in {various data analysis papers
\cite{
Damour:Iyer:Sathyaprakash:2000,
Damour:Iyer:Sathyaprakash:2001,
Damour:Iyer:Sathyaprakash:2001:err,
Arun:Iyer:Sathyaprakash:Sundararajan:2005}
}.
% For a future publication that includes higher-order RR formulas, the binary inspiral has been worked out to 3PN in
% references \cite{Arun:Blanchet:Iyer:Qusailah:2008}, \cite{Arun:Blanchet:Iyer:Qusailah:2008-1} and \cite{Arun:Blanchet:Iyer:Sinha:2009}.

\subsection{Considerations about limiting of the series}
Infinite series have to be restricted to finite ones for
practical issues. In paper 1 we gave instructions how to
limit the 1PN series as we set up the following properties
concerning the orders of {involved} terms:
\begin{enumerate}
	\item Bessel functions of order $n$: $J_n(n \, e_t) \sim {\cal O}(e_t^n)$,
		\label{Item::OrderBF}
	\item The $v(u)$ expansion coefficients: $\vKE{n}(n e_t) \sim {\cal O}(e_t^n) $,
	\item Even-in-$u$ expansion coefficients: $ \AKE{n}{j} \sim {\cal O}(e_t^{j})$,
	\item Odd-in-$u$ expansion coefficients: $\SAKE{n}{j} \sim {\cal O}(e_t^{j-1})$,
	\item The $\sin v$ expansion: $J'_n(e_t)=\frac{1}{2}\left(J_{n-1}(e_t)-J_{n+1}(e_t)\right) = {\cal O}(e_t^{n-1})$,
	\item Double series expansions: $\ProdCS{j} \sim {\cal O}(e_t^j) $ and $\ProdSS{j} \sim {\cal O}(e_t^{j-1}) $.
\end{enumerate}
These computations are still valid at 2PN and have to be applied to
each series where $e_t^{\rm sum~ index}$ plays a role. The result
of this is obvious but lengthy, so we skip the provision.
The interested reader may take a look at \cite{Watson:1980}
% and \cite{Colwell:1993}
for more information on an estimate of the error
when using finite sums.
Having double, triple, \dots , maximally $n$-tuples of summations,
each evaluated up to some order ${\cal O}( e_t^\Max)$ and, thus, containing $\Max$
terms (plus or minus some finite number), we have a computational cost of
$\sim \Max^n$ terms per each time step $\Mean_k$ and
${\rm NSP} \times \Max^n$ term computations {in total}, where NSP
is the number of sampling points, $k=(0,\dots, {\rm NSP}-1)$.
In our case, {a typical value could be $n=5$}.
%=======================================================
\section{Conclusion}

In this article we provide 2PN
accurate GW forms in terms of tensor spherical harmonics.
The analytic Fourier-Bessel series of the inverted KE
as well as the sines and cosines of the eccentric anomaly
have been evaluated. These series may be interesting
for perturbation theory of celestial mechanics. We could
verify parts of the results of \cite{Junker:Schafer:1992}
and reproduce results of the ideas of paper~1.

One {slight} drawback should be mentioned.
Without appropriate optimizing, the presented routine
is far less than quickly computable, for example for a
data rate of 4096 points per second and several minutes
to be observed and a restriction to errors of, say,
$<{\cal O}(e_t^5)$. A first numerical insight, done
in C
% by Jan Sperrhake
for the 1PN case, showed a
CPU time of $\sim$ one minute for the case of 128 data
points at an error of ${\cal O}(e_t^6)$.
It can, for example, give an impression of the orders
of magnitude of how many harmonics may have to be included
for a data analysis investigation.
Its CPU time consumption should be improved by atomising
the series computation to make it attractive for researchers in data analysis.
% and is, without investment
% of advance of computation, only few more than of academic
% interest.

A future investigation may include the 3PN GW amplitude
\cite{Arun:Blanchet:Iyer:Sinha:2009} and the 3PN QKP for point particles
without spin, or even more the spin dependent multipole moments
of \cite{	Faye:Blanchet:Buonanno:2006,
		Blanchet:Buonanno:Faye:2006,
		Blanchet:Buonanno:Faye:2006:err,
		Blanchet:Buonanno:Faye:2006:err:2}
and \cite{Porto:Ross:Rothstein:2010}
using the QKP for aligned spinning compact binaries and a well-suited and optimized
numerical implementation.
%=======================================================
\begin{acknowledgement}
We thank Jan Sperrhake for useful discussions. MT likes to thank
Johannes Hartung for an enlightening introduction on automatise and
efficient programming of {\em Mathematica}, in which we coded most
of our work.
This work was funded by the Deutsche Forschungsgemeinschaft
(DFG) through SFB/TR7 ``Gravitationswellenastronomie'' and
the DLR through ``LISA -- Germany''.
\end{acknowledgement}
%=======================================================
\appendix
\section{The solution to the 2PN accurate Kepler Equation:
an alternative way to derive it and a useful check}
The 2PN accurate KE,
\label{Sec::App_KeplerEQ}
\begin{eqnarray}
\label{Eq::App::M2PN}
 \Mean	&=&	u-e_t \sin u + \epsilon^4 \, \left( {\cal F}_{v-u} (v-u) +{\cal F}_v \sin v \right)
\nonumber \\
&=:&
u-e_t \sin u + \epsilon^4 \, {\cal F}_{4} (u)
\,,
\end{eqnarray}
can be inverted by defining
\begin{eqnarray}
 g_{} \left(\Mean - \epsilon^4 \, {\cal F}_{4} (u) \right)
=u%_{\rm 2PN}
\,.
\end{eqnarray}
as the appropriate solution, however it will look like.
We can Taylor expand it around the Newtonian solution,
\begin{eqnarray}
\label{Def::g2PN}
 {g_{\rm N} }\left(\Mean) - g'_N (\Mean) \, \epsilon^4 \, {\cal F}_{4} (u) \right)
=u%_{\rm 2PN}
\,.
\end{eqnarray}
The Newtonian solution $g$ is known,
\begin{eqnarray}
 g_{\rm N} \left( \Mean \right) = \Mean + \sum_{n=1}^{\infty} \frac{2}{n} J_n(n\,e_t)\, \sin n\Mean\,,
\end{eqnarray}
and (\ref{Def::g2PN}) reads
\begin{eqnarray}
\label{Eq::AppU2PNApprox}
 u%_{\rm 2PN} 
(\Mean)
&=&
%\left(
%\Mean + \sum_{n=1}^{\infty} \frac{2}{n} J_n(n\,e_t)\, \sin n\Mean
g_N(\Mean)
%\right)
-\left( 1 + \sum_{n=1}^{\infty}2J_{n}(n\,e_t)\,\cos{n\Mean}\right)\epsilon^4 {\cal F}_{4}\left(g_N(\Mean)\right)
\,.
\end{eqnarray}
We know
\begin{eqnarray}
\label{Eq::AppVMU}
 v-u	&=& \sum_{n=1}^{\infty}\left[\vKE{n}(n\,e_t)
	 - \frac{2}{n}J_n(n\,e_t) \right]\,\sin{n\Mean}\,, \\
\label{Eq::AppSinV}
 \sin v	&=& \frac{\sqrt{1-e_t^2} \, \sin u}{1-e_t\, \cos u} =
\sqrt{1-e_t^2} \, 2 \sum_{n=1}^{\infty} J_n'(n\,e_t)\,\sin{n\Mean}\,.
\end{eqnarray}
Inserting (\ref{Eq::AppVMU}) and (\ref{Eq::AppSinV})
into (\ref{Eq::AppU2PNApprox}) and applying the product rule
for a $\sin$ and a $\cos$ series, we obtain the result
of (\ref{Eq::KE2PN_analytic}).
This is a nice calculation, so let us show it in detail.

We abbreviate
\begin{equation}
 {\cal F}_{4}(g_{\rm N}(\Mean)) = \sum_{n=1}^{\infty} \alpha_n \sin{n\Mean}\,,
\end{equation}
and read the coefficients $\alpha$ from Equations (\ref{Eq::AppVMU}) and (\ref{Eq::AppSinV}).
In fact, for the comparison, their form does not matter.
Equation (\ref{Eq::AppU2PNApprox}) together with the rule for products of
a sin and a cos series reads
\begin{eqnarray}
 u &=& \sum_{n=1}^{\infty} \frac{J_n (n e_t)}{n} \sin n\Mean
\nonumber \\ &&
-\epsilon^4 \sum_{n=1}^{\infty} \left(\alpha_n + 2\ProdCS{n} [J_k (k e_t); \alpha_k ]\right) \sin{n\Mean}\,,\\
2 \ProdCS{n} [J_k (k e_t); \alpha_k ] 
&\stackrel{1}{=}& \left\{
\Theta(n,1)
 \sum_{k=1}^{n-1} \alpha_{n-k} J_k (k e_t)
+\sum_{k=1}^{\infty} \alpha_{n+k} J_k (k e_t)
-\sum_{k=n+1}^{\infty} \alpha_{k-n} J_k (k e_t)
\right\}
\nonumber \\
&\stackrel{2}{=}&
 \sum_{k=1}^{n-1}	\alpha_k J_{n-k}((n-k)e_t)
+\sum_{k=n+1}^{\infty}	\alpha_k J_{n-k}((n-k)e_t)
\nonumber \\ &&
-\sum_{k=1}^{\infty}	\alpha_k J_{n+k}((n+k)e_t)
\,.
\end{eqnarray}
$\Theta$ just dropped out as for $n=1$, $\alpha_k$
for $k \le 0$ vanish anyway.
In summation with $\alpha_k$ the last line gives
$\sum_{k=1}^{\infty}\alpha_k \left(J_{n-k}((n-k)e_t) {-} J_{n+k}((n+k)e_t) \right)$,
remembering
that for the $k=n$ term, $J_0 (0)=1$.

Let us, in contrast, directly derive the expansion coefficients via integration
and assume that $ g$ is the solution to the KE,
$u={g}(\Mean)$. Then, at $n \pi$ ($n \in \mathbb Z$), there are fixed points of the KE:
$n\pi = u = \Mean$ and ${  g}(\Mean) - \Mean$ can be expressed in $\sin$ series,
\begin{eqnarray}
{  g}(\Mean) - \Mean &=& \sum_{n=1}^{\infty} A_n \, \sin (n\Mean)\,.
\end{eqnarray}
The expansion coefficients, directly computed via
integration read
\begin{eqnarray}
 A_n &\stackrel{1}{=}&
\frac{2}{\pi}\int_{0}^{\pi} \left[ { g}(\Mean)-\Mean \right] \sin(n\Mean){\rm d}\Mean
\nonumber \\
&\stackrel{2}{=}&
-\frac{2}{n\pi}\int_{0}^{\pi} \left[ { g}(\Mean)-\Mean \right] {\rm d} (\cos n\Mean)
\nonumber\\
&\stackrel{3}{=}&
[{\rm boundary=0}]+\frac{2}{n\pi} \int_{0}^{\pi} \cos(n\Mean)[{ g}'(\Mean)-1]{\rm d}\Mean
\nonumber \\
%             A_n	
&\stackrel{4}{=}&	\frac{2}{n\pi}\int_{0}^{\pi}
\cos \left\{ n\Mean \right\} {\rm d}({ g}(\Mean))
\nonumber \\
&\stackrel{5}{=}&	\frac{2}{n\pi}\int_{0}^{\pi}
\cos \left\{ n\,\left( u-e_t\, \sin u + \epsilon^4 {\cal F}_{4}(u) \right) \right\}
{\rm d}u
\nonumber \\
&\stackrel{6}{=}& \frac{2}{n\pi}\int_{0}^{\pi}
\cos \left\{ n\,\left( u-e_t\, \sin u \right) \right\}
{\rm d}u
- \frac{2}{\pi}\int_{0}^{\pi}
\sin \left\{ n\,\left( u-e_t\, \sin u \right) \right\}
\epsilon^4 \sum_{k=1}^{\infty} \alpha_k \sin{k\Mean}
\nonumber \\
&\stackrel{7}{=}&
\frac{2 J_n(ne_t)}{n} - \epsilon^4
\frac{1}{\pi}\int_{0}^{\pi} \sum_{k=1}^{\infty}
\alpha_k
\left(
  \cos \left[ {(k-n)(u-e_t \sin u)}  \right]
- \cos \left[ {(k+n)(u-e_t \sin u)}  \right]
\right)
{\rm d}u
\nonumber \\
&\stackrel{8}{=}&
\frac{2 J_n(ne_t)}{n} - \epsilon^4
\sum_{k=1}^{\infty}
\alpha_k
{
\left(
  J_{k-n}(n e_t)
- J_{k+n}(n e_t)
\right)
}\,,
\end{eqnarray}
and we see that the 2PN coefficient shows agreement in both calculations.
In step 5 we used Equation (\ref{Eq::App::M2PN}), in step 6 we Taylor expanded
the argument of $\cos$ around the Newtonian $\Mean$ and in step 7 we
used that only Newtonian $\Mean$ is required in the sum.
The trigonometrics  of $u$ are dealt with equivalently.

\section{Fourier representation: $\sin{mu}$ and $\cos{mu}$ at 2PN}
\label{Appendix::Cos_mu}
As an exemplary calculation, we determine the
expansion coefficients of $\cos {mu}$,
\begin{equation}
% \sin {mu}	&=&	\sum_{j=1}^{\infty} \SKE{m}{j} \sin {j\Mean} \,, \\
 \cos {mu}	=	\sum_{j=0}^{\infty} \CKE{m}{j} \cos {j\Mean} \,.
\end{equation}
Using integration by parts, the computation turns out to be
\begin{eqnarray}
\CKE{m}{j}
&=&	\frac{2}{\pi} \int_{0}^{\pi} \cos{mu} \cos{j\Mean}\, {\rm d}\Mean
\nonumber \\
&=& [\cos{mu} \sin{j\Mean} ]_{0}^{\pi}- \frac{2}{j\pi} \int_{0}^{\pi} ({\rm d}\cos{mu})\, \sin{j\Mean} \, {\rm d}\Mean
\nonumber \\
&=& [\dots]_{0}^{\pi}- \frac{2m}{j\pi} \int_{0}^{\pi} (\sin{mu})\, \sin{j\Mean}
\,\frac{{\rm d} u}{{\rm d}\Mean}{\rm d}\Mean
\nonumber \\
&=& \frac{m}{j\pi} \int_{0}^{\pi}
\left(\cos[mu+j\Mean_{\rm 2PN}] - \cos[mu-j\Mean_{\rm 2PN}] \right) 
\,{\rm d}u
\nonumber \\
&=& \frac{m}{j\pi} \int_{0}^{\pi}
\left(\cos[mu+j\Mean_{\rm N}] - \cos[mu-j\Mean_{\rm N}] \right) 
\,{\rm d}u
\nonumber \\ &&
-\frac{m}{\pi} \int_{0}^{\pi}
\left(
\sin[mu+j\Mean_{\rm N}] - \sin[mu-j\Mean_{\rm N}]
\right)
\,
\epsilon^4 \sum_{f=1}^{\infty} \alpha_f \sin{f\Mean_{\rm N}}
\,{\rm d}u
\nonumber \\
&=&
\frac{m}{j}\left( J_{j+m}(j e_t) - J_{j-m}(j e_t) \right)
\nonumber \\
&&-\frac{m}{2 \pi}
\sum_{n=1}^{\infty}
\int_{0}^{\pi} \alpha _n
\Bigl[
\cos (j \Mean-m u-\Mean n)
-\cos (j \Mean-m u+ \Mean n)
\nonumber \\ &&
-\cos (j \Mean+m u-\Mean n)
+\cos (j \Mean+m u+\Mean n)
\Bigr]
{\rm d}u
\nonumber \\
 &=&
\frac{m}{j}\left( J_{j+m}(j e_t) - J_{j-m}(j e_t) \right)
\nonumber \\
&&
-\sum_{n=1}^{\infty} \frac{m \alpha_n}
{{2 \pi }}
\int_{0}^{2\pi}
\Bigl\{ \nonumber \\ &&
\quad \cos \left[(j-n)\left(u-e_t \sin (u)\right)-m u\right] \nonumber \\ &&
-\cos \left[(j+n) \left(u-e_t \sin (u)\right)-m u\right] \nonumber \\ &&
-\cos \left[(j-n) \left(u-e_t \sin (u)\right)+m u\right] \nonumber \\ &&
+\cos \left[(j+n) \left(u-e_t \sin (u)\right)+m u\right]
\Bigr\}{\rm d}u
\,.
\end{eqnarray}
The task is now to bring these integrals to the form
\begin{equation}
\frac{1}{\pi}\int_{0}^{\pi} \cos\left[ {\bf x}_i  (u- {\bf y}_i \,\sin u) \right]
=J_{{\bf x}_i} ( {\bf x}_i  \, {\bf y}_i )\,,
\end{equation}
with some prefactor ${\bf x}$ and ``eccentricity'' ${\bf y}$ to be determined for each special case.
In order of appearance in the last equation above, these
 eccentricities read
\begin{eqnarray}
{\bf y}_{1}  &= \frac{e_t (j-n)}{j-m-n} \,, \\
{\bf y}_{2}  &= \frac{e_t (j+n)}{j-m+n} \,, \\
{\bf y}_{3}  &= \frac{e_t (j-n)}{j+m-n} \,, \\
{\bf y}_{4}  &= \frac{e_t (j+n)}{j+m+n} \,,
\end{eqnarray}
${\bf x}_i$ is simply the denominator.
Herewith, the rest is easy to calculate and so are the coefficients of $\sin mu$.
%=======================================================
\section{The relevant tensor spherical harmonics}
\label{Sec::TensorSphericalHarmonics}
We take the definitions from \cite{Turner:Will:1978}, Equations (A1) -- (A5) therein,
\begin{eqnarray}
 T^{(m)}_{\rm LM}	&=&	A_{LM} \left( \hat{\theta} \hat{\theta} -\hat{\phi} \hat{\phi} \right)	-	i\,B_{LM} \left( \hat{\theta} \hat{\phi} +\hat{\phi} \hat{\theta} \right)		\,,\\
 T^{(e)}_{\rm LM}		&=&	B_{LM} \left( \hat{\theta} \hat{\theta} -\hat{\phi} \hat{\phi} \right)	-	i\,A_{LM} \left( \hat{\theta} \hat{\phi} +\hat{\phi} \hat{\theta} \right)		\,,\\
A_{LM}				&=&	 2\,C_L \, \Biggl\{	\frac{\partial^2}{\partial \, \theta^2}	+ \frac{L(L+1)}{2}	\Biggr\} \, Y_{LM} (\theta,\phi) \,, \\
B_{LM}				&=&	-2\,C_L \, \Biggl\{	\frac{\partial}{\partial \theta}	- \cot{\theta}	\Biggr\} \, Y_{LM} (\theta,\phi) \,, \\
C_L					&=&	\left[ 2\,L\,(L+1)(L+2)(L-1) \right]^{-1/2}\,, \\
T^{E2}_{LM}			&=&	T^{(m)}_{LM} \,,\\
T^{B2}_{LM}			&=&	-i\,T^{(e)}_{LM} \,,
\end{eqnarray}
with $\hat \theta$ and $\hat \phi$ being basis unit vectors in $\theta$ and $\phi$ direction ($\theta$ is the angle between
the orbital angular momentum and the line--of--sight vector {\bf N}, and $\phi$ measures the angle from the x axis to {\bf N}
projected onto the (x, y) plane, see Figure 3 of \cite{Turner:Will:1978}).
The relevant multipoles read
\begin{eqnarray}
% L=2
T^{E2}_{22}			&=&	\frac{1}{8} \sqrt{\frac{5}{2 \pi }} e^{2 i \phi } \left[{\TensorProdttpp} \frac{1}{2} (\cos (2 \theta )+3)	
									+	 2 i \cos (\theta ) {\TensorProdtppt}  \right]		\,, \\
T^{E2}_{21}			&=&	\frac{1}{4} \sqrt{\frac{5}{2 \pi }} e^{2 i \phi } \left[
						\cos (\theta ) \sin (\theta ) \TensorProdttpp
					+	i \sin (\theta )\TensorProdtppt
						\right]
						\,,\\
T^{E2}_{20}			&=&	\frac{1}{8}	\sqrt{\frac{15}{\pi }} \sin ^2(\theta )\TensorProdttpp \,,\\ 
T^{B2}_{22}			&=&	\frac{-1}{16}	\sqrt{\frac{5}{2 \pi }} e^{2 i \phi } 	\left[ {\TensorProdtppt} (\cos (2 \theta )+3)-4 i {\TensorProdttpp} \cos (\theta ) 	\right] \,,\\
T^{B2}_{21}			&=&	\frac{1}{4}	\sqrt{\frac{5}{2 \pi }}i e^{i \phi } \left[{\TensorProdttpp}+i {\TensorProdtppt} \cos (\theta ) 					\right] \sin (\theta ) \,,\\ 
T^{B2}_{20}			&=&	\frac{-1}{8}	\sqrt{\frac{15}{\pi }} {\TensorProdtppt} \sin ^2(\theta ) \,,\\ 
% L=3
T^{E2}_{33}			&=&	-\frac{1}{32}	\sqrt{\frac{21}{\pi }}  e^{3 i \phi } \left[ 4 i {\TensorProdtppt} \cos (\theta )+{\TensorProdttpp} (\cos (2 \theta )+3) \right] \sin (\theta ) \,,\\ 
T^{E2}_{32}			&=&	\frac{1}{32} 	\sqrt{\frac{7}{2 \pi }} e^{2 i \phi } \left[ 8 i {\TensorProdtppt} \cos (2 \theta )+{\TensorProdttpp} (5 \cos (\theta )+3 \cos (3 \theta ))\right] \,,\\ 
T^{E2}_{31}			&=&	\frac{1}{32} 	\sqrt{\frac{35}{\pi }}  e^{i \phi }  \left[ 4 i {\TensorProdtppt} \cos (\theta )+{\TensorProdttpp} (3 \cos (2 \theta )+1)\right]  \sin (\theta )\,,\\
T^{E2}_{30}			&=&	\frac{1}{8}	\sqrt{\frac{105}{\pi }} {\TensorProdttpp} \cos (\theta ) \sin ^2(\theta )		\,, \\
 T^{B2}_{33}			&=&	\frac{1}{32}	\sqrt{\frac{21}{\pi }}	e^{3 i \phi }  \left[ {\TensorProdtppt} (\cos (2 \theta )+3)-4 i {\TensorProdttpp} \cos (\theta ) \right] \sin (\theta )	\,, \\
 T^{B2}_{32}			&=&	\frac{-1}{32}	\sqrt{\frac{7}{2 \pi }}	e^{2 i \phi } \left[ {\TensorProdtppt} (5 \cos (\theta )+3 \cos (3 \theta ))-8 i {\TensorProdttpp} \cos (2  \theta ) \right]		\,, \\
 T^{B2}_{31}			&=&	\frac{-1}{32}	\sqrt{\frac{35}{\pi }}	e^{i \phi } \left[ {\TensorProdtppt} (3 \cos (2 \theta )+1)-4 i {\TensorProdttpp} \cos (\theta ) \right] \sin (\theta ) \,, \\
 T^{B2}_{30}			&=&	\frac{-1}{8}	\sqrt{\frac{105}{\pi }} {\TensorProdtppt} \cos (\theta ) \sin ^2(\theta )		\,, \\
%
% L=4 =======================================================================
\hline
 T^{E2}_{44}			&=&	\frac{3}{32}	\sqrt{\frac{7}{2 \pi }} e^{4 i \phi } \left[ 4 i {\TensorProdtppt} \cos (\theta )+{\TensorProdttpp} (\cos (2 \theta )+3) \right] \sin ^2(\theta   )	\,, \\
 T^{E2}_{43}			&=&	\frac{-3}{32}	\sqrt{\frac{7}{\pi }} e^{3 i \phi } \left[4 {\TensorProdttpp} \cos ^3(\theta )+i {\TensorProdtppt} (3 \cos (2 \theta )+1)\right] \sin (\theta ) \,, \\
 T^{E2}_{42}			&=&	\frac{3}{64	\sqrt{2 \pi }} e^{2 i \phi } \Bigl[ 2 i {\TensorProdtppt} (\cos (\theta )+7 \cos (3 \theta ))
\nonumber \\ &&
+{\TensorProdttpp} (4 \cos (2 \theta )+7 \cos (4 \theta )+5) \Bigr]  \,, \\
 T^{E2}_{41}			&=&	\frac{3}{32 	\sqrt{\pi }} e^{i \phi } \left[ i {\TensorProdtppt} (7 \cos (2 \theta )+5)+{\TensorProdttpp} (5 \cos (\theta )+7 \cos (3 \theta )) \right] \sin (\theta ) \,, \\
 T^{E2}_{40}			&=&	\frac{3}{32}	\sqrt{\frac{5}{\pi }} {\TensorProdttpp} (7 \cos (2 \theta )+5) \sin ^2(\theta )	\,, \\
% %
 T^{B2}_{44}			&=&	\frac{-3}{32}	\sqrt{\frac{7}{2 \pi }} e^{4 i \phi }  \left[ {\TensorProdtppt} (\cos (2 \theta )+3)-4 i {\TensorProdttpp} \cos (\theta ) \right] \sin ^2(\theta ) \,, \\
 T^{B2}_{43}			&=& \frac{3}{32} 	\sqrt{\frac{7}{\pi }} e^{3 i \phi }  \left[ 4 {\TensorProdtppt} \cos ^3(\theta )-i {\TensorProdttpp} (3 \cos (2 \theta )+1) \right] \sin (\theta )	\,, \\
 T^{B2}_{42}			&=&	\frac{-3}{64	\sqrt{2 \pi }} e^{2 i \phi } \Bigl[{\TensorProdtppt} (4 \cos (2 \theta )+7 \cos (4 \theta )+5)
\nonumber \\ &&
-2 i {\TensorProdttpp} (\cos (\theta )+7 \cos (3 \theta )) \Bigr] \,, \\
 T^{B2}_{41}			&=&	\frac{-3}{32	\sqrt{\pi }} e^{i \phi } \left[ {\TensorProdtppt} (5 \cos (\theta )+7 \cos (3 \theta ))-i {\TensorProdttpp} (7 \cos (2 \theta )+5) \right] \sin (\theta ) \,, \\
 T^{B2}_{40}			&=&	\frac{-3}{32}	\sqrt{\frac{5}{\pi }} {\TensorProdtppt} (7 \cos (2 \theta )+5) \sin ^2(\theta )	\,, \\
%
% L=5 =======================================================================
\hline
 T^{E2}_{55}			&=&	-\frac{1}{64}	\sqrt{\frac{165}{\pi }} e^{5 i \phi }  \left[ 4 i {\TensorProdtppt} \cos (\theta )+{\TensorProdttpp} (\cos (2 \theta )+3) \right] \sin ^3(\theta ) \,, \\
 T^{E2}_{54}			&=&	\frac{1}{64}	\sqrt{\frac{33}{2 \pi }}  e^{4 i \phi } \Bigl[ 8 i {\TensorProdtppt} (2 \cos (2 \theta )+1)
\nonumber \\ &&
+{\TensorProdttpp} (19 \cos (\theta )+5 \cos (3 \theta )) \Bigr] \sin ^2(\theta )	\,, \\
 T^{E2}_{53}			&=&	-\frac{1}{256}	\sqrt{\frac{33}{\pi }} e^{3 i \phi }  \Bigl[4 i {\TensorProdtppt} (7 \cos (\theta )+9 \cos (3 \theta ))
\nonumber \\
					& &+{\TensorProdttpp} (28 \cos (2 \theta )+15 \cos (4 \theta )+21) \Bigr] \sin  (\theta )	\,, \\
 T^{E2}_{52}			&=&	\frac{1}{128}	\sqrt{\frac{11}{2 \pi }} e^{2 i \phi }  \Bigl[ 8 i {\TensorProdtppt} (\cos (2 \theta )+3 \cos (4 \theta ))
\nonumber \\
					&  &+{\TensorProdttpp} (14 \cos (\theta )+3 (\cos (3 \theta )+5 \cos (5  \theta ))) \Bigr]	\,, \\
 T^{E2}_{51}			&=&	\frac{1}{128}	\sqrt{\frac{77}{2 \pi }} e^{i \phi }  \Bigl[4 i {\TensorProdtppt} (5 \cos (\theta )+3 \cos (3 \theta ))
\nonumber \\
					&& +{\TensorProdttpp} (12 \cos (2 \theta )+15 \cos (4 \theta )+5) \Bigr] \sin  (\theta )	\,, \\
 T^{E2}_{50}			&=&	\frac{1}{64}	\sqrt{\frac{1155}{\pi }} {\TensorProdttpp} \left[ 5 \cos (\theta )+3 \cos (3 \theta ) \right] \sin ^2(\theta )	\,, \\
 T^{B2}_{55}			&=&	\frac{1}{64}	\sqrt{\frac{165}{\pi }} e^{5 i \phi }  	\left[ {\TensorProdtppt} (\cos (2 \theta )+3)-4 i {\TensorProdttpp} \cos (\theta )	\right]  \sin ^3(\theta )	\,, \\
 T^{B2}_{54}			&=&	-\frac{1}{64}	\sqrt{\frac{33}{2 \pi }} e^{4 i \phi }  \Bigl[ {\TensorProdtppt} (19 \cos (\theta )+5 \cos (3 \theta ))
\nonumber \\ 
				&& -8 i {\TensorProdttpp} (2 \cos (2 \theta )+1) \Bigr] \sin ^2(\theta )	\,, \\
 T^{B2}_{53}			&=&	\frac{1}{256}	\sqrt{\frac{33}{\pi }} e^{3 i \phi }  	\Bigl[ {\TensorProdtppt} (28 \cos (2 \theta )+15 \cos (4 \theta )+21)
\nonumber \\
					&&	-4 i {\TensorProdttpp} (7 \cos (\theta )+9 \cos (3 \theta )) \Bigr] \sin  (\theta )	\,, \\
 T^{B2}_{52}			&=&	 -\frac{1}{128}	\sqrt{\frac{11}{2 \pi }} e^{2 i \phi }  \Bigl[ {\TensorProdtppt} (14 \cos (\theta )+3 (\cos (3 \theta )+5 \cos (5 \theta )))
\nonumber \\
					&&	-8 i {\TensorProdttpp} (\cos (2 \theta )+3 \cos (4  \theta )) \Bigr] \,, \\
 T^{B2}_{51}			&=&	-\frac{1}{128}	\sqrt{\frac{77}{2 \pi }} e^{i \phi }  	\Bigl[ {\TensorProdtppt} (12 \cos (2 \theta )+15 \cos (4 \theta )+5)
\nonumber \\
					&&	-4 i {\TensorProdttpp} (5 \cos (\theta )+3 \cos (3 \theta )) \Bigr] \sin  (\theta )	\,, \\
 T^{B2}_{50}			&=&	-\frac{1}{64}	\sqrt{\frac{1155}{\pi }} {\TensorProdtppt} \left[5 \cos (\theta )+3 \cos (3 \theta ) \right] \sin ^2(\theta )	\,, \\
\hline
%
% L=6 =======================================================================
 T^{E2}_{6 6}			&=&	\frac{3}{256}	\sqrt{\frac{715}{2 \pi }} e^{6 i \phi } 	\left[ 4 i {\TensorProdtppt} \cos (\theta )+{\TensorProdttpp} (\cos (2 \theta )+3) \right] \sin ^4(\theta )	\,, \\
 T^{E2}_{6 5}			&=&	-\frac{1}{256}	\sqrt{\frac{2145}{2 \pi }} e^{5 i \phi } 	\Bigl[ 2 i {\TensorProdtppt} (5 \cos (2 \theta )+3)
\nonumber \\
					&& +{\TensorProdttpp} (13 \cos (\theta )+3 \cos (3 \theta )) \Bigr] \sin ^3(\theta )	\,, \\
 T^{E2}_{6 4}			&=&	\frac{1}{1024}	\sqrt{\frac{195}{\pi }} e^{4 i \phi }  \Bigl[ 8 i {\TensorProdtppt} (13 \cos (\theta )+11 \cos (3 \theta ))
\nonumber \\
					&& +{\TensorProdttpp} (92 \cos (2 \theta )+33 \cos (4 \theta )+67) \Bigr] \sin ^2(\theta)	\,, \\
 T^{E2}_{6 3}			&=&	-\frac{3}{1024}	\sqrt{\frac{13}{2 \pi }} e^{3 i \phi }  \Bigl[2 i {\TensorProdtppt} (52 \cos (2 \theta )+55 \cos (4 \theta )+21)
\nonumber \\
					&&	+{\TensorProdttpp} (126 \cos (\theta )+75 \cos (3 \theta )+55 \cos (5  \theta )) \Bigr] \sin (\theta )	\,, \\
 T^{E2}_{6 2}			&=&	\frac{1}{4096}	\sqrt{\frac{13}{2 \pi }}e^{2 i \phi }  \Bigl[ 4 i {\TensorProdtppt} (10 \cos (\theta )+81 \cos (3 \theta )+165 \cos (5 \theta ))
\nonumber \\
					&&	+{\TensorProdttpp} (289 \cos (2 \theta )+30 (\cos (4  \theta )+7)+495 \cos (6 \theta )) \Bigr] 	\,, \\
 T^{E2}_{6 1}			&=&	\frac{1}{1024}	\sqrt{\frac{65}{\pi }} e^{i \phi }  \Bigl[ 2 i {\TensorProdtppt} (60 \cos (2 \theta )+33 \cos (4 \theta )+35)
\nonumber \\
					&&	+{\TensorProdttpp} (70 \cos (\theta )+87 \cos (3 \theta )+99 \cos (5 \theta   )) \Bigr]  \sin (\theta )	\,, \\
 T^{E2}_{6 0}			&=&	\frac{1}{512} \sqrt{\frac{1365}{2 \pi }} {\TensorProdttpp} \left[ 60 \cos (2 \theta )+33 \cos (4 \theta )+35 \right] \sin ^2(\theta )	\,.
\end{eqnarray}
%=======================================================
\section{The multipole components in terms of Fourier coefficients}
\label{Appendix::Results}

%=======================================================
%
%
% bibliography
\providecommand{\WileyBibTextsc}{}
\let\textsc\WileyBibTextsc
\providecommand{\othercit}{}
\providecommand{\jr}[1]{#1}
\providecommand{\etal}{~et~al.}

\bibliographystyle{adp}
%\bibliographystyle{utphys} % arXiv 
% \bibliography{../references}
%\begin{thebibliography}{100}
\providecommand{\WileyBibTextsc}{}
\let\textsc\WileyBibTextsc
\providecommand{\othercit}{}
\providecommand{\jr}[1]{#1}
\providecommand{\etal}{~et~al.}

%\end{thebibliography}
\end{document}